\documentclass[twoside,leqno,twocolumn]{article}

\usepackage[letterpaper]{geometry}

\usepackage{geometry}
\usepackage{ltexpprt}
\usepackage{hyperref}

\usepackage{xr}
\makeatletter
\newcommand*{\addFileDependency}[1]{
  \typeout{(#1)}
  \@addtofilelist{#1}
  \IfFileExists{#1}{}{\typeout{No file #1.}}
}
\makeatother

\newcommand*{\myexternaldocument}[1]{
    \externaldocument{#1}
    \addFileDependency{#1.tex}
    \addFileDependency{#1.aux}
}
\myexternaldocument{a_SDM_SI}

\usepackage{amsmath,amsfonts,bm} % for math commands
\usepackage{booktabs} % for tables
\usepackage{graphicx}
\usepackage{caption}
\usepackage[justification=centering]{subcaption}
\usepackage{wrapfig}
\usepackage{tabularx,ragged2e}
\usepackage{multirow} % added by Yixuan
\usepackage[algo2e,ruled,vlined]{algorithm2e} %
\newcolumntype{C}{>{\Centering\arraybackslash}X} % centered "X" column, used to make automatic line breaks in tables

\newcommand{\bb}{\hspace{-1mm} $\bullet$}

\usepackage{bbm} % added by Yixuan for indicator funciton

\newcommand{\SSS}{\textsc{SSSNET}}

\usepackage{cleveref} 

\newcommand{\ER}{Erd\H{o}s-R\'enyi}

\usepackage{xcolor}
\newcommand{\red}[1]{\textcolor{red}{\textbf{#1}}}
\newcommand{\blue}[1]{\textcolor{blue}{\underline{#1}}}

\begin{document}

\title{\Large SSSNET: Semi-Supervised Signed Network Clustering}
\author{Yixuan He\thanks{yixuan.he@stats.ox.ac.uk; Univ. of Oxford (UoOx).}
\and Gesine Reinert\thanks{reinert@stats.ox.ac.uk; UoOx \& The Alan Turing Inst. (ATI).}
\and Songchao Wang\thanks{wscwdy@mail.ustc.edu.cn; Univ. of Sci. and Tech. of China.}
\and Mihai Cucuringu\thanks{mihai.cucuringu@stats.ox.ac.uk; UoOx \& ATI.}}

\date{}

\maketitle

% Copyright Statement
% When submitting your final paper to a SIAM proceedings, it is requested that you include
% the appropriate copyright in the footer of the paper.  The copyright added should be
% consistent with the copyright selected on the copyright form submitted with the paper.
% Please note that "20XX" should be changed to the year of the meeting.

% Default Copyright Statement
\fancyfoot[R]{\scriptsize{Copyright \textcopyright\ 2022 by SIAM\\
Unauthorized reproduction of this article is prohibited}}

% Depending on which copyright you agree to when you sign the copyright form, the copyright
% can be changed to one of the following after commenting out the default copyright statement
% above.

%\fancyfoot[R]{\scriptsize{Copyright \textcopyright\ 20XX\\
%Copyright for this paper is retained by authors}}

%\fancyfoot[R]{\scriptsize{Copyright \textcopyright\ 20XX\\
%Copyright retained by principal author's organization}}

%\pagenumbering{arabic}
%\setcounter{page}{1}%Leave this line commented out.

\begin{abstract} \small\baselineskip=9pt 
Node embeddings are a powerful tool in the analysis of networks; yet, their full potential for the important task of node clustering has not been fully exploited. In particular, most state-of-the-art methods 
generating node embeddings of \textit{signed} networks focus on link sign prediction, and those that pertain to node clustering are usually not graph neural network (GNN) methods. Here, we introduce a novel probabilistic balanced normalized cut loss for training nodes in a GNN framework for semi-supervised signed network clustering, called {\SSS}. 
The method is end-to-end in combining embedding generation and clustering without an intermediate step; it has node clustering as main focus, with an emphasis on polarization effects arising in networks. 
The main novelty of our approach is a new take on the role of social balance theory for signed network embeddings. The standard heuristic for justifying the criteria for the embeddings hinges on the assumption that an ``enemy's enemy is a friend". Here, instead, a neutral stance is assumed on whether or not the enemy of an enemy is a friend. 
Experimental results on various data sets, including a synthetic signed stochastic block model, a polarized version of it, and real-world data at different scales,
demonstrate that {\SSS} can achieve comparable or better results than state-of-the-art spectral clustering methods, for a wide range of noise and sparsity levels.
{\SSS} complements existing methods through the possibility of including exogenous information, in the form of node-level features or labels.

\small\baselineskip=9pt\noindent\textbf{Keywords}{: clustering, 
signed networks,
signed stochastic block models, 
graph neural networks, 
generalized social balance, 
polarization.}
\end{abstract}

\vspace{-3mm}
\section{Introduction}
In social network analysis, 
signed network clustering is an important task which standard community detection algorithms do not address  \cite{su2021comprehensive}. 
Signs of edges in networks may indicate positive or negative sentiments, see for example
\cite{sampson1969novitiate}. 
As an illustration, review websites as well as online news allow users to approve or denounce others \cite{leskovec2010signed}. 
Clustering time series can be viewed as an instance of signed network clustering \cite{aghabozorgi2015time}, 
with the empirical correlation matrix being construed as a weighted signed network.
For recommendation systems,
\cite{RecSignedSocialNet_WWW16} introduced a principled approach to capturing local and global information from signed social networks mathematically, and proposed a novel recommendation framework.
Recently interest has grown on the topic of polarization in social media, mainly fueled by a large variety of speeches and statements made in the pursuit of public good, and their impact on the integrity of democratic processes \cite{Xiao}; our work also contributes to the growing literature of polarization in signed networks.

Most competitive state-of-the-art methods  
generating node embeddings for signed networks
focus on link sign prediction \cite{Javari,Chen18,Xu,Lee,Chen,yan2021muse, Derr}, and those that pertain to node clustering are 
not GNN methods \cite{Wang17, Xu,kunegis2010spectral,Chiang,SPONGE_AISTATS_2019}. Here, 
we introduce a graph neural network (GNN) framework, called {\SSS}, with a 
\textit{Signed Mixed-Path Aggregation} (SIMPA) scheme, to obtain node embeddings for signed clustering.

The main novelty of our approach is a new take on the role of social balance theory for signed network embeddings. The standard 
heuristic for justifying the criteria for the embeddings \cite{
Xu, chen2018, Derr, li2020learning, Huang2019,huang2101sdgnn} hinges on 
social balance theory \cite{harary1953notion,sharma2021balance}, or \emph{multiplicative distrust propagation} as in \cite{guha2004propagation}, which asserts that in a social network, in a triangle either all three nodes are friends, or two friends have a common enemy; otherwise it would be viewed as {\it unbalanced}. More generally, all cycles are assumed to prefer to contain either zero or an even number of negative edges. This hypothesis 
is difficult to justify for general signed networks. 
For example, the enemies of enemies are not necessarily friends, see \cite{guha2004propagation, tang2014distrust}; an example is given by
the social network of relations between 16 tribes in
New Guinea 
\cite{read1954cultures}.
Hence, the  present work takes a neutral stance on whether or not the enemy of an enemy is a friend, thus generalizing the
\emph{atomic propagation} by \cite{guha2004propagation}. 
\begin{figure*}[ht]
\centering
\vspace{-10mm}
\includegraphics[width=\linewidth]{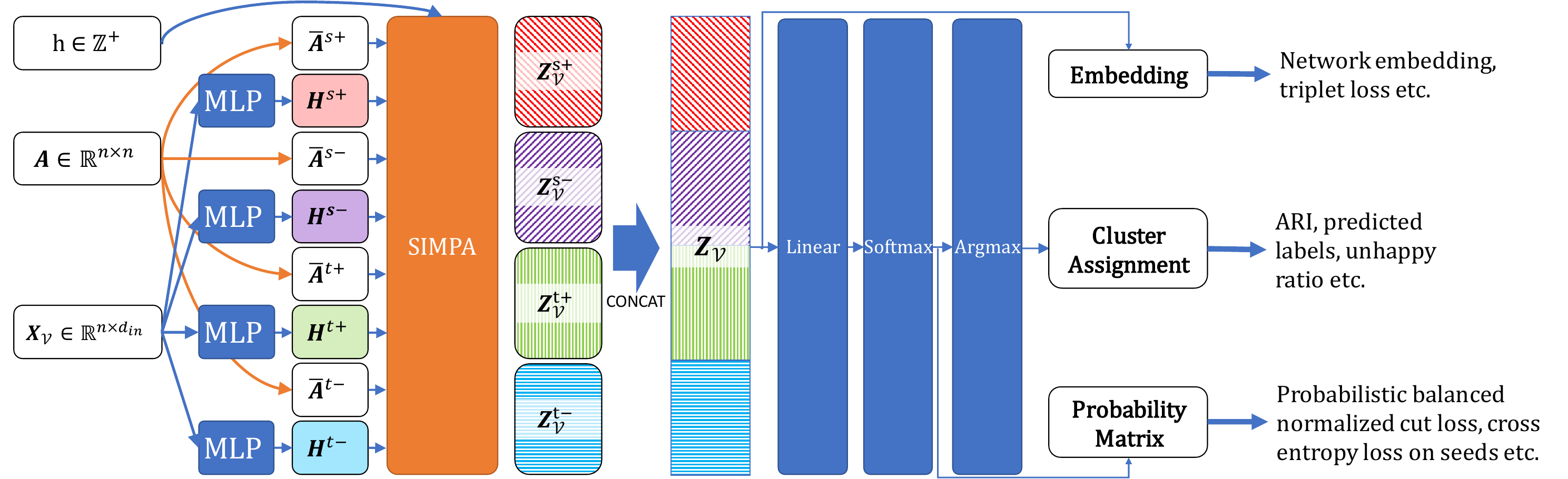}
\vspace{-7.5mm}
\caption{
{\SSS} overview: starting from feature matrix $\mathbf{X}_\mathcal{V}$ and adjacency matrix $\mathbf{A}$, we first compute the row-normalized adjacency matrices $\overline{\mathbf{A}}^{s+}, \overline{\mathbf{A}}^{s-}, \overline{\mathbf{A}}^{t+}$, 
$\overline{\mathbf{A}}^{t-}.$ 
We then apply four separate MLPs on $\mathbf{X}_\mathcal{V},$ to obtain hidden representations
$\mathbf{H}^{s+},\mathbf{H}^{s-},\mathbf{H}^{t+},$
$\mathbf{H}^{t-},$ respectively. 
Next, we compute their decoupled embeddings via Eq.~(\ref{SSSNET_eq:g_sp}) and its equivalent for negative/target embeddings. 
The concatenated decoupled embeddings are the final embeddings. We add another linear layer followed by a unit softmax function to obtain the probability matrix $\mathbf{P}$. Applying argmax to each row of $\mathbf{P}$ yields cluster assignments for all nodes.}
\label{SSSNET_fig:method_overview}
\end{figure*}

From a method's viewpoint, the neutral stance is reflected in the feature aggregation in SIMPA, which provides the basis of the network embedding. Network clustering is then carried out using the node embedding as input and a loss function for training; see Figure \ref{SSSNET_fig:method_overview} for an overview.
To train {\SSS},
the loss function consists of a self-supervised 
novel probabilistic balanced  normalized cut loss acting on all training nodes, and a supervised loss which acts on seed nodes, if available. Experimental results at different scales demonstrate that our method 
achieves  
state-of-the-art performance on synthetic data, for a wide range of network densities, while complementing other methods through the possibility of incorporating exogenous information. Tested on real-world data for which ground truth is available,
our method outperforms its competitors in terms of  the 
Adjusted Rand Index \cite{hubert1985comparing}.

\textbf{Main contributions.}  Our main contributions are  as follows:
\bb~(1) An efficient end-to-end 
GNN for semi-supervised signed node clustering, 
based on a new variant of social balance theory. 
To the best of our knowledge, this is the first GNN-based method deriving node embeddings for clustering signed networks, potentially with attributes. 
The advantage of the ability to handle  features is that we can incorporate eigen-features of other methods (such as eigenvectors of the signed Laplacian),  thus borrowing strength from existing methods.
\bb~(2) 
A \textit{Signed Mixed-Path Aggregation} (SIMPA) framework based
on our new take on social balance theory. 
\bb~(3) A
probabilistic version of balanced normalized cut to serve as a self-supervised loss function for signed clustering. 
\bb~(4) We achieve state-of-the-art performance on various signed clustering tasks, including a challenging version of a classification task, for which we customize and adapt a general definition of polarized signed stochastic block models (\textsc{Pol-SSBM}), to include an ambient cluster and multiple polarized SSBM communities, not necessarily of equal size, but of equal density. 
Code and preprocessed data are available at \url{https://github.com/SherylHYX/SSSNET_Signed_Clustering}. An alternative implementation is available in the open-source library \textit{PyTorch Geometric Signed Directed} \cite{he2022pytorch}. 
%%%%%%%%%%%%%%%%%%%%%%%%%%%%%%%%%%%%%%%%%%%%

%%%%%%%%%%%%%%%%%%%%%%%%%%%%%%%%%%%%%%%%%%
\vspace{-2.5mm}
\section{Related Work}  \label{SSSNET_sec:related}

\vspace{-1mm}
\subsection{Network Embedding and Clustering}
We introduce a semi-supervised method for node embeddings, which uses the idea of an aggregator and relies on powers of adjacency matrices for such aggregators. 
The use of an aggregation function is motivated by  
\cite{Hamilton2017}. 
The use of powers of adjacency matrices for neighborhood information aggregation was sparked by a  mechanism for node feature aggregation, proposed in \cite{Tian}, with a data-driven similarity metric during training. 

Non-GNN methods have been employed for signed clustering. \cite{kunegis2010spectral} uses the signed Laplacian matrix and its normalized versions for signed clustering.  
\cite{mercado2016clustering} clusters signed networks using
the geometric mean of Laplacians.In
\cite{Chiang}, nodes are clustered based on
optimizing the Balanced Normalized Cut and the Balanced Ratio Cut. 
\cite{zheng2015spectral} develops two normalized signed Laplacians based on the so-called SNScut and BNScut.
\cite{SPONGE_AISTATS_2019} relies on a generalized eigenproblem and achieves state-of-the-art performance on signed clustering. 
To conduct semi-supervised structural learning on signed networks, \cite{LIU2020105714} uses variational Bayesian inference and \cite{cucuringu2019mbo} devises an MBO scheme. \cite{Bhowmick2019} tackles network sparsity.  
However, these prior works do not take node attributes into account. 
\cite{Wang17} exploits the network structure and
node attributes simultaneously, but does not utilize known labels. 
\cite{Xu} views relationship formation between users as the comprehensive effects of latent factors and trust transfer patterns, via social balance theory.

GNNs 
have also been used for signed network embedding tasks, but not for signed clustering. 
SGCN \cite{Derr} employs social balance theory to aggregate and propagate the information across layers. Considering joint interactions between positive and negative edges
is another main inspiration for our method.
Many other GNNs \cite{ huang2101sdgnn, Huang2019,li2020learning,Chen18,Lee,yan2021muse} are also based on social balance theory, usually applied to data with strong positive class imbalance.
Numerous other signed network embedding methods 
\cite{Chen,chen2018,wang2017signed,Javari,Xu} also do not explore the node clustering problem. 
Hence we do not employ these methods for comparison.
%%%%%%%%%%%%%%%%%%%%%%%%%%%%%%%%%%%

\vspace{-3mm}
\subsection{Polarization}
Opinion formation in a social network can be driven by a few small groups which have a polarized opinion, in a social network in which many agents do not (yet) have a strongly formed opinion. These small clusters could strongly influence the public discourse and even threaten the integrity of democratic processes. Detecting such small clusters of polarized agents in a network of ambient nodes is hence of interest \cite{Xiao}. 
For a network with node set $\mathcal{V}$, \cite{Xiao} introduces the notion of a polarized community structure $(\mathcal{C}_1,\mathcal{C}_2)$ within the wider network, as two disjoint sets of nodes $\mathcal{C}_1,\mathcal{C}_2\subseteq \mathcal{V},$ such that
\bb 
(1) there are relatively few (resp. many) negative (resp. positive) edges within $\mathcal{C}_1$ and within $\mathcal{C}_2;$
\bb 
(2) there are relatively few (resp. many) positive (resp. negative) edges across $\mathcal{C}_1$ and $\mathcal{C}_2;$
\bb
(3)  there are relatively few edges (of either sign) from $\mathcal{C}_1$ and $\mathcal{C}_2$ to the rest of the graph.
By allowing a subset of nodes to be neutral with respect to the polarized 
structure, \cite{tzeng2020discovering} derives a  
formulation in which each cluster inside a polarized community  
is naturally characterized by the solution to the maximum discrete Rayleigh's quotient (MAX-DRQ) problem. However, this model cannot incorporate node attributes. 
Extending the approach in \cite{bonchi2019discovering}  and \cite{SPONGE_AISTATS_2019}, here, a community structure $(\mathcal{C}_1,\mathcal{C}_2)$  is said to be {\it polarized}   
if (1) and (2) hold, while (3) is not required to hold.  Moreover, our model includes different parameters for the noise and edge probability.

%%%%%%%%%%%%%%%%%%%%%%%%%%%%%%%%%%%%%%%%%%%%%%%%%

%%%%%%%%%%%%%%%%%%%%%%%%%%%%%%%%%%%%%%%%%%%%%%%%%%

\vspace{-3mm}
\section{The {\SSS} Method}  \label{SSSNET_sec:proposed}
\subsection{Problem Definition}
Denote a signed 
network with node attributes as $\mathcal{G}=(\mathcal{V}, \mathcal{E}, w, \mathbf{X}_\mathcal{V}),$ where $\mathcal{V}$ is the set of nodes, $\mathcal{E}$ is the set of (directed) edges or links, and $w \in (-\infty, \infty)^{| \mathcal{E}|}$ is the set of weights of the edges. 
Here $\mathcal{G}$ could have self-loops but not multiple edges. 
The total number of nodes is $n=|\mathcal{V}|,$ and
$\mathbf{X}_\mathcal{V}\in \mathbb{R}^{n\times d_\text{in}}$ is a matrix whose rows are node attributes. These attributes could be generated from the adjacency matrix $\mathbf{A}$ with entries
$\mathbf{A}_{ij}=w_{ij}$, the edge weight, if there is an edge
between nodes $v_i$ and $v_j$; otherwise  $\mathbf{A}_{ij}=0$.
We decompose 
$\mathbf{A}$ into its positive and negative part $\mathbf{A}^+$ and $\mathbf{A}^-,$ where
$\mathbf{A}_{ij}^+ = \max(\mathbf{A}_{ij},0)$ and $\mathbf{A}_{ij}^- = -\min(\mathbf{A}_{ij},0).$ 
A clustering into $K$ clusters is 
a partition of the node set  
into disjoint sets
$\mathcal{V}=\mathcal{C}_0\cup\mathcal{C}_1\cup\cdots\cup\mathcal{C}_{K-1}.$ 
Intuitively, nodes within a cluster should be similar to each other, while nodes across clusters should be dissimilar. In a semi-supervised setting, for each of the $K$ clusters, a fraction of training nodes are selected as seed nodes, for which 
the cluster membership labels are known before training. 
The set of seed nodes is denoted as $\mathcal{V}^\text{seed}\subseteq\mathcal{V}^\text{train}\subset\mathcal{V},$
where $\mathcal{V}^\text{train}$ is the set of all training nodes.
For this task, 
the goal is to use the  embedding for assigning each node $v \in \mathcal{V}$ to a cluster containing known seed nodes. 
When no seeds are given, we are in a self-supervised setting, where only the number of clusters, $K,$ is given.
%%%%%%%%%%%%%%%%%%%%%%%%%%%%%%%%
\vspace{-3mm}
\subsection{Path-Based Node Relationship}
Methods based on social balance theory assume that, given a negative relationship between $v_1, v_2$ and a negative relationship between $v_2, v_3$, the nodes
$v_1$ and $v_3$ should be positively related.
This assumption may be sensible for social networks, but in other networks such as correlation networks \cite{aghabozorgi2015time, arfaoui2017oil}, it is not obvious why it should hold.
Indeed, the column $|\Delta^\text{u}|$ in Table \ref{SSSNET_tab:data_sets} counts the number of triangles with an odd number of negative edges in eight real-world data sets and one synthetic model.
We observe that $|\Delta^\text{u}|$ is never zero, and that in some cases, the percentage of unbalanced  triangles (the last column) is quite large, such as in Sampson's network of novices and the simulated SSBM($n=5000, K=5, p=0.1, \rho=1.5$) (``Syn" in Table \ref{SSSNET_tab:data_sets}, SSBM is defined in Section \ref{subsecssbm}). 
The relative high proportion of unbalanced triangles sparks our novel approach. 
{\SSS} holds a neutral attitude towards the relationship between $v_1$ and $v_3.$
In contrast to social balance theory, 
our definition of ``friends'' and ``enemies'' is based on the set of paths within a given length between any two nodes.
For a target node $v_j$ to be an $h$-hop ``friend" neighbor of source node $v_i$ \emph{along a given path} from $v_i$ to $v_j$ of length $h$, all edges on this path need to be positive. For a target node $v_j$  to be an ${h}$-hop ``enemy" neighbor of source node $v_i$ along a given path from $v_i$ to $v_j$ of length $h$, exactly one edge on this path has to be negative. Otherwise, $v_i$ and $v_j$ are neutral to each other on this path. For directed networks, only directed paths are taken into account, and the friendship relationship is no longer symmetric. 

\begin{figure}[ht] 
\centering
\includegraphics[width=\linewidth]{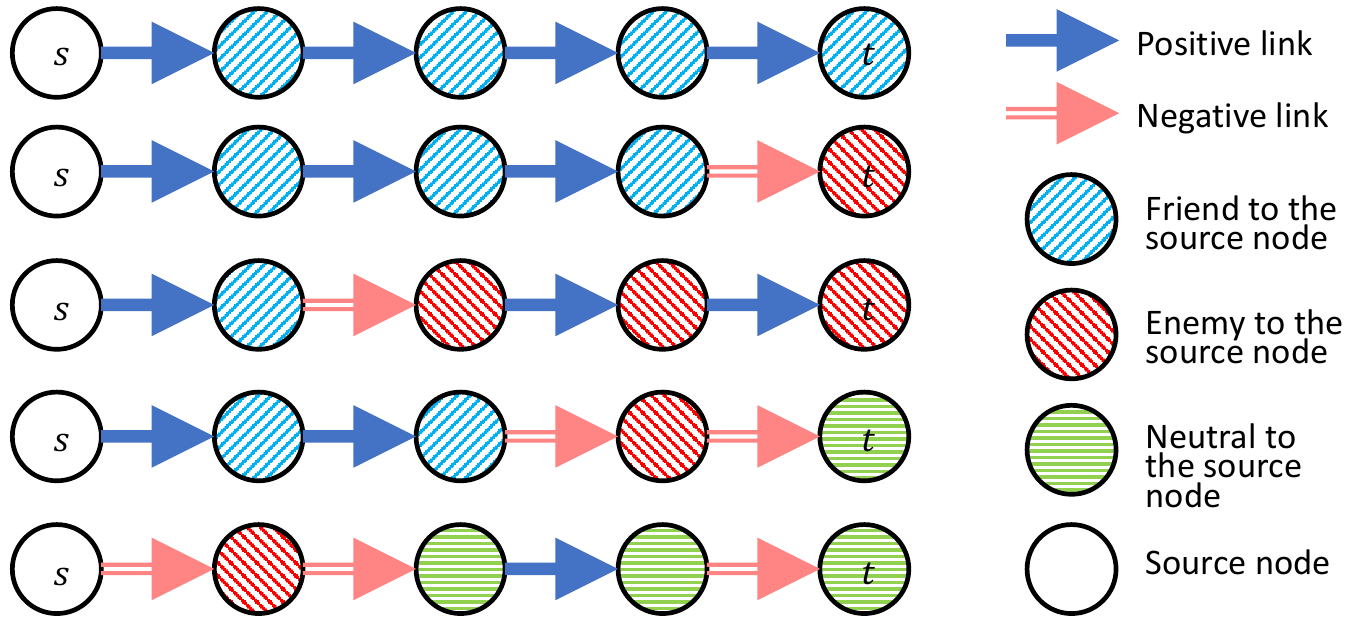}
\vspace{-7mm}     
\caption{Example: five paths between the source (s) and target (t) nodes, and resulting relationships. While we assume a neutral relationship on the last two paths, social balance theory claims them as "friend" and "enemy", respectively.}
    \label{SSSNET_fig:relationship_paths}
\end{figure}

Figure \ref{SSSNET_fig:relationship_paths} illustrates five different paths of length four, connecting the source and the target nodes. 
We can also obtain the relationship of a source node to a target node within a path by reversing the arrows in Figure \ref{SSSNET_fig:relationship_paths}.
Note that it is possible for a node to be both a ``friend" and an ``enemy" to a source node simultaneously, as there might be multiple paths between them, with different resulting relationships. 
Our model aggregates these relationships by assigning different weights to different paths connecting two nodes.    
For example, the source node and target node may have all five paths shown in Figure \ref{SSSNET_fig:relationship_paths} connecting  them. Since the last two paths are neutral paths and do not cast a vote on their relationship, we only take the top three paths into account.  
We refer to the long-range neighbors whose information would be considered by a node as the \emph{contributing neighbors} with respect to the node of interest. 
\vspace{-8mm}
%%%%%%%%%%%%%%%%%%%%%%%%%%%%%%%
\subsection{Signed Mixed-Path Aggregation (SIMPA)}
\label{SSSNET_sec:SIMPA}
SIMPA aggregates
neighbor information from contributing neighbors within $h$ hops, 
by a weighted average of the embeddings of the up-to-$h$-hop contributing neighbors of a node, with the weights constructed in analogy to a random walk on the set of nodes.
%%%%%%%%%%%%%%%%%%%%%%%%%%%%%%%%%%
\vspace{-4mm}
\subsubsection{SIMPA Matrices}
\label{sec:SIMPA}
\emph{First}, we row-normalize 
$\mathbf{A}^+$ and $\mathbf{A}^-$, 
to obtain matrices  $\mathbf{\overline{A}}^{s+}$ and $\mathbf{\overline{A}}^{s-}$, respectively. 
Inspired by the regularization discussed in \cite{kipf2016semi}, 
we add a weighted self-loop to each node and carry out 
the normalization by setting $\mathbf{\overline{A}^{s+} = (\Tilde{D}}^{s+})^{-1}\mathbf{\Tilde{A}}^{s+},$ where $\mathbf{\Tilde{A}}^{s+} = \mathbf{A}^+ + \tau^+\mathbf{I}$ and the diagonal matrix $\mathbf{\Tilde{D}}^{s+}(i,i) = \sum_j \mathbf{\Tilde{A}}^{s+}(i,j),$ for some $\tau^+\geq 0$; 
similarly, we row-normalize  $\mathbf{A}^{s-}$ to obtain $\mathbf{\overline{A}}^{s-}$ based on $\tau^-.$
\emph{Next}, we explore multi-hop neighbors by taking powers or mixed powers of $\mathbf{\overline{A}}^{s+}$ and $\mathbf{\overline{A}}^{s-}.$
The $h$-hop ``friend" neighborhood can be computed directly from $(\overline{\mathbf{A}}^{s+})^h$,  the $h^{th}$ power of $\overline{\mathbf{A}}^{s+}$. 
Similarly, the $h$-hop ``enemy" neighborhood is computed directly from the mixed powers of $h-1$ terms of $\overline{\mathbf{A}}^{s+},$ and exactly one term of $\mathbf{\overline{A}}^{s-}.$ 
We denote the set of \emph{up-to-$h$-hop} ``friend" neighborhood matrices as 
$\mathcal{A}^{s+,h}=\{(\mathbf{\overline{A}}^{s+})^{h_1}:h_1\in\{0,\cdots,h\}\},$ where $(\mathbf{\overline{A}}^{s+})^0=\mathbf{I},$ the identity matrix, 
and the set of up-to-$h$-hop ``enemy" neighborhood matrices as 
$$
\vspace{-1mm} 
\mathcal{A}^{s-,h}=
     \left\{(\mathbf{\overline{A}}^{s+})^{h_1}\cdot\mathbf{\overline{A}}^{s-}\cdot(\mathbf{\overline{A}}^{s+})^{h_2}:h_1,h_2\in H 
   \right\} \vspace{-1mm} 
$$
\noindent with 
$H=\{(h_1,h_2): h_1, h_2\in\{0,\ldots,h-1\},h_1+h_2\leq h-1\}$. 
With added self-loops, any $h$-hop neighbor defined by our matrices \textit{aggregates beliefs} 
from nearby neighbors. 
Since a node 
is not 
an enemy to itself, we set $\tau^-=0.$ We use $\tau^+=\tau=0.5$ in our experiments.  

When the signed network is directed,
we additionally carry out 
$\ell_1$
row normalization and calculate mixed powers for 
$(\mathbf{A}^+)^T$ and $(\mathbf{A}^-)^T.$
We denote the row-normalized adjacency matrices for \emph{target} positive and 
negative as $\overline{\mathbf{A}}^{t+}$ and $\overline{\mathbf{A}}^{t-},$ respectively. 
Likewise, we denote the set of up-to-$h$-hop target ``friend" (resp. ``enemy") neighborhood matrices as $\mathcal{A}^{t+,h}$ (resp. $\mathcal{A}^{t-,h}$).
%%%%%%%%%%%%%%%%%%%%%%%%%%%%%%%%%%%
\subsubsection{Feature Aggregation Based on SIMPA}

\vspace{-3mm}
Next, we define four feature-mapping functions for source positive, source negative, target positive and target negative embeddings, respectively. 
A source positive embedding of a node is 
the weighted combination of its contributing neighbors' hidden representations, for neighbors up to $h$ hops away.
The source positive hidden representation is denoted as $\mathbf{H}_\mathcal{V}^{s+}\in \mathbb{R}^{n\times d}$. 
Assume
that 
each node in $\mathcal{V}$ has a vector of features 
and summarize these features in the input feature matrix 
$\mathbf{X_\mathcal{V}}$. 
The source positive embedding $\mathbf{Z}^{s+}_\mathcal{V}$ is given by 
\vspace{-2mm} 
\begin{equation}
\vspace{-2mm} 
    \mathbf{Z}^{s+}_\mathcal{V}=
    %\left(
    \sum_{\mathbf{M}\in \mathcal{A}^{s+,h}} \omega_{{\mathbf{M}}}^{s+} \cdot {\mathbf{M}}
    %\right) 
    \cdot \mathbf{H}_\mathcal{V}^{s+} \in \mathbb{R}^{n\times d},
    \label{SSSNET_eq:g_sp}
\end{equation}
where 
for each $\mathbf{M},$ $\omega_{\mathbf{M}}^{s+}$ is a learnable scalar, and $d$ is the embedding dimension.  
In our experiments, we use $\mathbf{H}_\mathcal{V}^{s+}=\textbf{MLP}^{(s+,l)}(\mathbf{X}_\mathcal{V}).$
The hyperparameter $l$ controls the number of layers in the multilayer perceptron (MLP) with ReLU activation;
we fix $l=2$  throughout. 
Each layer of the MLP has the same number $d$ of hidden units.

The embeddings
$\mathbf{Z}^{s-}_\mathcal{V}$,   $\mathbf{Z}^{t+}_\mathcal{V}$ and $\mathbf{Z}^{t-}_\mathcal{V}$
for source negative embedding, target positive embedding and target negative embedding, respectively, are defined similarly. 
Different parameters for the MLPs for different embeddings are possible. 
We concatenate the embeddings to obtain the final node embedding  
as a $n \times (4d)$ matrix
$
\mathbf{Z}_\mathcal{V}=\operatorname{CONCAT}\left(\mathbf{Z}_\mathcal{V}^{s+}, \mathbf{Z}_\mathcal{V}^{s-},\mathbf{Z}_\mathcal{V}^{t+},\mathbf{Z}_\mathcal{V}^{t-}\right) 
.$
\noindent The  embedding vector $\mathbf{z}_i$ for a 
node $v_i,$ 
is the $i^\text{th}$ row of $\mathbf{Z}_\mathcal{V}$, namely 
$ 
\mathbf{z}_i:=(\mathbf{Z}_\mathcal{V})_{(i,:)} \in \mathbb{R}^{4d}.$ 
Next we apply a linear layer to $\mathbf{Z}_\mathcal{V}$ so that the resulting matrix has
the same number of columns as the number $K$ of clusters. 
We apply the unit \textit{softmax} function to map each row to a probability vector $\mathbf{p}_i\in \mathbb{R}^{K}$ of length equal to the number of clusters,  with entries denoting the probabilities of each node to belong to each cluster. The resulting probability matrix is denoted as $\mathbf{P}\in \mathbb{R}^{n\times K}.$
If the input network is undirected, it suffices to find
$\mathcal{A}^{s+,h}$
and $\mathcal{A}^{s-,h},$ and we obtain the final embedding as
$\mathbf{Z}_\mathcal{V}=\operatorname{CONCAT}\left(\mathbf{Z}_\mathcal{V}^{s+}, \mathbf{Z}_\mathcal{V}^{s-}\right) \in \mathbb{R}^{n\times (2d)}.$  

%%%%%%%%%%%%%%%%%%%%%%%%%%%%%%%%
\vspace{-3mm}
\subsection{Loss, Overview \& Complexity Analysis}

Node clustering is optimized to minimize a loss function which pushes 
embeddings of nodes within the same cluster close to each other, while driving apart embeddings of nodes from different clusters.
We first introduce a novel self-supervised loss function for node clustering, then discuss supervised loss functions when labels are available for some seed nodes.
 
\vspace{-3mm}
\subsubsection{Probabilistic Balanced Normalized Cut Loss}
\label{SSSNET_sec:pbnc}
For a  clustering $(\mathcal{C}_0, \ldots \mathcal{C}_{K-1})$, let $\{\mathbf{x}_0,\cdots,\mathbf{x}_{K-1}\}$ denote the cluster indicator vectors so that $\mathbf{x}_{k}(i)=1 $ if node $i$ is in cluster $C_k$, and 0 otherwise. Let $\mathbf{L}^+=\mathbf{D}^+-\mathbf{A}^+$ denote the unnormalized graph Laplacian for the positive part of $\mathbf{A},$ where $\mathbf{D}^+$ is a diagonal matrix whose diagonal entries are row-sums of $\mathbf{A}^+.$
Then $\mathbf{x}_k^T\mathbf{L}^+\mathbf{x}_k$ measures the total weight of positive edges linking cluster ${\mathcal{C}_k}$ to \emph{other} clusters. 
Further, $\mathbf{x}_k^T\mathbf{A}^-\mathbf{x}_k$ measures the total weight of negative edges \emph{within} cluster ${\mathcal{C}_k}$. 
Since $\mathbf{D}^+-\mathbf{A}
=\mathbf{D}^+-\mathbf{A}^++\mathbf{A}^-$, 
then 
$\mathbf{x}_k^T(\mathbf{L}^++\mathbf{A}^-)\mathbf{x}_k=\mathbf{x}_k^T(\mathbf{D}^+-\mathbf{A})\mathbf{x}_k$ measures the total weight of the \textit{unhappy} edges with respect to cluster ${\mathcal{C}_k}$; 
``unhappy edges" 
violate their expected signs (positive edges across clusters or negative edges within clusters). 
The loss function in this paper is related to the (non-differentiable)  Balanced Normalized Cut (BNC) \cite{Chiang}.
 In analogy, 
we introduce the differentiable \textit{Probabilistic Balanced Normalized Cut (PBNC) loss}
\vspace{-2mm}
\begin{equation}
    \mathcal{L}_\text{PBNC} = 
    \sum_{k=1}^K\frac{(\mathbf{P}_{(:,k)})^T(\mathbf{D}^+-\mathbf{A})\mathbf{P}_{(:,k)}}{(\mathbf{P}_{(:,k)})^T\mathbf{\overline{D}}\mathbf{P}_{(:,k)}},  
    \label{SSSNET_eq:loss_pbnc}
    \vspace{-2mm} 
\end{equation}
where $\mathbf{P}_{(:,k)}$ denotes the $k^{th}$ column of the probability matrix $\mathbf{P}$ 
and  $\overline{\mathbf{D}}_{ii} = \sum_{j=1}^n |A_{ij}| $. As 
column $k$ of $\mathbf{P}$ is a relaxed version of $\mathbf{x}_k$, 
the numerator in Eq.~(\ref{SSSNET_eq:loss_pbnc}) is a probabilistic count of the number of unhappy edges.

\vspace{-3mm}
%%%%%%%%%%%%%%%%%%%%%%%%%%%%%%%%
\subsubsection{Supervised Loss}
When some seed nodes have known labels, a supervised loss can be added to the loss function. 
For nodes in $\mathcal{V}^\text{seed},$ we use as a supervised loss function similar to that in \cite{Tian}, the sum of a cross entropy loss $\mathcal{L}_\text{CE}$ and a triplet loss. The triplet loss is
\begin{equation}
\mathcal{L}_\text{triplet} = \frac{1}{|\mathcal{T}|}\sum_{(v_i,v_j,v_k) \in \mathcal{T}}
\hspace{-3mm} 
ReLU(\operatorname{CS}(\mathbf{z}_i,\mathbf{z}_j)-\operatorname{CS}(\mathbf{z}_i,\mathbf{z}_k)+\alpha),
\label{SSSNET_eq:L_triplet}
\vspace{-2mm}
\end{equation}
\noindent 
where $\mathcal{T}\subseteq \mathcal{V}^\text{seed}\times \mathcal{V}^\text{seed}\times \mathcal{V}^\text{seed}$ is a set of node triplets: $v_i$ is an anchor seed node, and $v_j$ is a seed node from the same cluster as the anchor, while $v_k$ is from a different cluster. Here,  $\operatorname{CS}(\mathbf{z}_i,\mathbf{z}_j)$ is the cosine similarity of the  
embeddings of nodes $v_i$ and $v_j$, 
chosen 
so as to avoid sensitivity to the magnitude of the embeddings. $\alpha\geq 0$ is the contrastive margin as in \cite{Tian}. 
$\mathcal{L}_\text{CE} + \gamma_t\mathcal{L}_\text{triplet}$ forms the supervised part of the loss function for \SSS, for a suitable parameter $\gamma_t>0.$ 
%%%%%%%%%%%%%%%%%%%%%%%%%%%%%%%%%

\vspace{-3mm}
\subsubsection{Overall Objective Function and Framework Overview}  \label{SSSNET_sec:ovObjFunc}
By combining $\mathcal{L}_\text{CE}$, Eq.~(\ref{SSSNET_eq:loss_pbnc}), and Eq.~(\ref{SSSNET_eq:L_triplet}), the 
objective function    
minimizes  
\begin{equation}
\vspace{-2mm}
    \mathcal{L} = \mathcal{L}_\text{PBNC}+\gamma_s( \mathcal{L}_\text{CE}+\gamma_t\mathcal{L}_\text{triplet}), 
   \label{SSSNET_eq:loss_overall}
\end{equation}
where $\gamma_s,\gamma_t>0$ are weights for the supervised part of the loss and triplet loss, respectively.
The final embedding can then be used, for example, for node clustering.
A linear layer coupled with a unit softmax function turns the embedding  into a probability matrix. A node is assigned to the cluster for which its membership probability is highest. 
Figure \ref{SSSNET_fig:method_overview} gives an overview. 

%%%%%%%%%%%%%%%%%%%%%%%%%%%%%%%%%%%%%%%%%%%%%%%%%

\vspace{-3mm}
\subsubsection{Complexity Analysis}
The matrix operations 
in Eq.~(\ref{SSSNET_eq:g_sp}) appear to be computationally expensive and space unfriendly. However, {\SSS} resolves these concerns via a sparsity-aware implementation, detailed in Algorithm~\ref{algo:SIMPA}
in SI~\ref{appendix_subsec:simpa},
without explicitly calculating the sets of powers, 
maintaining sparsity throughout.
Therefore, for input feature dimension $d_\text{in}$ and hidden dimension $d$, if $d'=\max(d_\text{in},d) \ll n,$ time and space complexity of SIMPA, and implicitly {\SSS}, is  $\mathcal{O}(|\mathcal{E}|d'h^2+4nd'K)$ and $\mathcal{O}(4|\mathcal{E}|+10nd'+nK),$ respectively \cite{complexity}. 
For large networks, SIMPA is amenable to a more scalable version following  \cite{Fey/etal/2021}.
%%%%%%%%%%%%%%%%%%%%%%%%%%%%%%%%%%%%%%%%%%%%%%%%

%%%%%%%%%%%%%%%%%%%%%%%%%%%%%%%%%%%%%%%%%%%%%%%%%

\vspace{-3mm}
\section{Experiments} \label{SSSNET_sec:experimentss}

This section  describes the synthetic and real-world data sets used in this  study, and illustrates the efficacy of our method. 
When ground truth is available, performance is measured by the Adjusted Rand Index (ARI) \cite{hubert1985comparing}. 
When no labels are provided, we measure performance by the ratio of number of ``unhappy edges" to that of all edges.
Our self-supervised loss function is applied to the subgraph induced by all training nodes.
We do not report Normalized Mutual Information (NMI) \cite{pmlr-v32-romano14} performance in Figure \ref{SSSNET_fig:synthetic_compare} (but reported in SI~\ref{SSSNET_appendix_subsec:synthetic_more}
) as it has some shortcomings \cite{amelio2015normalized}, and results from the ARI and NMI from our synthetic experiments indeed yield almost the same ranking for the methods, with average Kendall tau 0.808 and standard deviation 0.233.

\subsection{Data}
\vspace{-3mm}
\subsubsection{Synthetic Data: Signed Stochastic Block Models (SSBM)} \label{subsecssbm} 

\vspace{-3mm}
A Signed Stochastic Block Model (SSBM) for a network on $n$ nodes  with $K$ blocks (clusters), is constructed  similar to \cite{SPONGE_AISTATS_2019} but with a more general cluster size definition. 

In our experiments, we choose the number of clusters, the (approximate) ratio, $\rho$, between the largest and the smallest cluster size, sign flip probability, $\eta$, and the number, $n$, of nodes. To tackle the hardest clustering task, all pairs of nodes within a cluster and those between clusters have the same edge probability, with more details in SI~\ref{SSSNET_appendix:SSBM_more}.
Our SSBM model can be represented by SSBM($n, K, p, \rho, \eta$).
%%%%%%%%%%%%%%%%%%%%%%%%%%%%%%%%%%
\vspace{-3mm}
\subsubsection{Synthetic Data: Polarized SSBMs}
In a polarized SSBM model, SSBMs are planted in an ambient network; each block of each SSBM is a cluster, and the nodes not assigned to any SSBM form an ambient cluster. 
The
polarized SSBM model
that creates communities of SSBMs, is generated as follows: 
\bb (1) Generate an {\ER} 
    graph with $n$ nodes and edge probability $p,$ whose sign is set to $ \pm 1$ with equal probability 0.5.
\bb   
(2) Fix $n_c$ as the number of SSBM communities,  and   calculate community sizes $N_1 \le N_2 \le \cdots \le N_{r},$ for each of the $r$ communities as in Section \ref{subsecssbm}, such that the ratio of the largest block size to the smallest block size is  
 approximately 
$\rho$, and the total number of nodes in these SSBMs is $N\times n_c.$ 
\bb 
(3) Generate $r$ SSBM models, each with $K_i=2, i = 1, \ldots,r$ blocks,  
number of nodes according to its community size, with the same edge probability $p$, size ratio $\rho$, and flip probability $\eta$. 
\bb
(4) Place the SSBM models on disjoint subsets of the whole network; the remaining nodes not part of any SSBM are dubbed
as {\it ambient} nodes.

\begin{figure}[!ht]
\centering
\vspace{-2mm}
\subcaptionbox[]{ Sorted adjacency matrix.
}[0.48\linewidth] 
{\includegraphics[width=\linewidth, trim=0cm 0cm 0cm 0.1cm,clip]{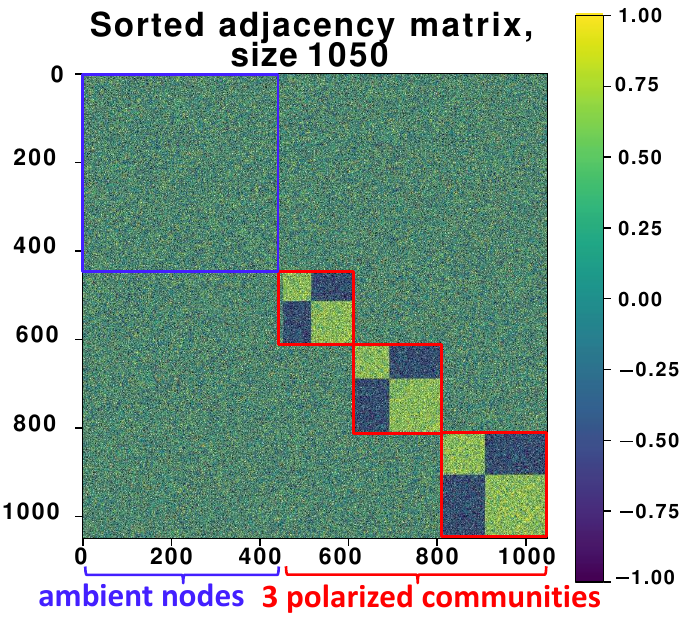}}
\subcaptionbox[]{Polarized community \#1.
}[0.5\linewidth]{\includegraphics[width=0.93\linewidth, trim=0cm 0cm 0cm 0.7cm,clip] {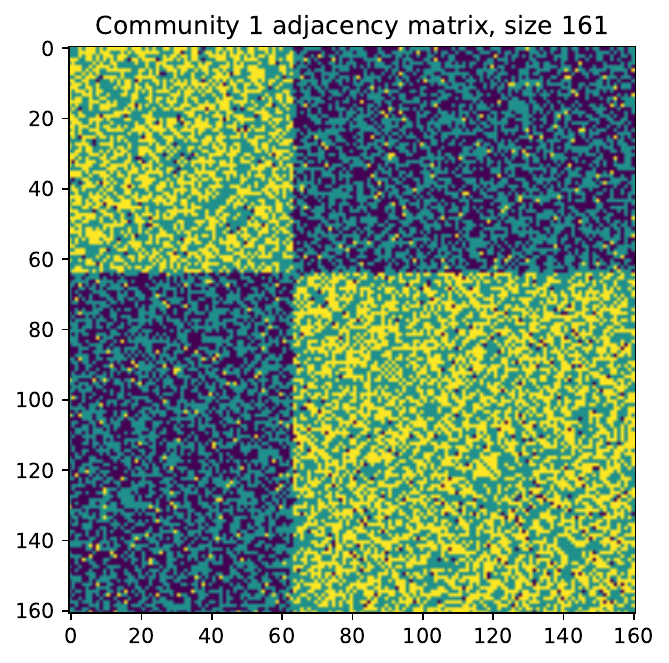} }
\vspace{-4mm} 
\caption{A polarized SSBM model with 
1050 nodes, $r=3$ polarized communities of sizes 161, 197, and 242; $\rho=1.5,$ default SSBM community size $N=200, p=0.5, \eta=0.05,$ and each SSBM has $K_1=K_2=K_3=2$ blocks, rendering $K=7$. 
} 
\label{SSSNET_fig:visualize_polarized}
\vspace{-5mm}
\end{figure}
Therefore, while the total number of clusters in an SSBM equals the number of blocks, the total number of clusters within a polarized SSBM model equals $K = 1+ \sum_{i=1}^{r} K_i = 1+2r.$ In our experiments, we also assume the existence of an ambient cluster. 
The resulting polarized SSBM model is denoted as {\textsc{Pol-SSBM}} ($n,r,p,\rho,\eta,N$).
The setting in \cite{bonchi2019discovering} can be construed as a special case of our model, see SI~\ref{SSSNET_appendix:polarized_discussion}.

Figure \ref{SSSNET_fig:visualize_polarized} gives a visualization of a polarized SSBM model with 1050 nodes, $p=0.5, \eta=0.05, N=200,$ with 3 SSBMs of $K=2$ blocks each, $\rho=1.5.$ 
The sorted adjacency matrix has its rows and columns sorted by cluster membership, starting with the ambient cluster. 
With $\rho=1.5,$ the largest SSBM community has size 242, while the smallest has size 161, as $\frac{242}{161}\approx 1.5=\rho.$
For $n=1050,$ the default size of a SSBM community is $N=200.$ For $n=5000$ (resp, $n=10000$) we consider $N=500$ (resp. $N=2000$).
%%%%%%%%%%%%%%%%%%%%%%%%%%%%%%%%%%%%%%%%%%%%

\vspace{-3mm}

\subsubsection{Real-World Data}
We perform experiments on six real-world signed network data sets (\textit{Sampson} \cite{sampson1969novitiate},
\textit{Rainfall} \cite{bertolacci2019climate},
\textit{Fin-YNet},
\textit{S\&P 1500} \cite{SP1500},
\textit{PPI} \cite{vinayagam2014integrating}, 
and \textit{Wiki-Rfa} \cite{west2014exploiting}), summarized in Table~\ref{SSSNET_tab:data_sets}.
 \textit{Sampson}, as the only data set with given node attributes (1D ``Cloisterville" binary attribute), cover four 
social relationships, which are combined into a network;  \textit{Rainfall} contains Australian rainfalls pairwise correlations; \textit{Fin-YNet} consists of 21 yearly financial correlation networks and its results are averaged; \textit{S\&P1500} is a correlation network 
of 
stocks during 2003-2005; \textit{PPI} is a signed protein-protein interaction network; \textit{Wiki-Rfa} describes voting information for electing Wikipedia managers.

We use labels given by each data set for \textit{Sampson} (5 clusters), and sector memberships for \textit{S\&P 1500} and \textit{Fin-YNet}  (10 clusters).
For \textit{Rainfall}, with 6 clusters, 
we use labels from SPONGE as proxy for ground truth to carry out semi-supervised learning.
For other data sets with no ``ground-truth" labels available, we train {\SSS} in a self-supervised manner, using all nodes to train. 
By exploring performance on SPONGE, we set the number of clusters for Wiki-Rfa as three, and similarly ten for PPI. 
Additional details concerning the data and preprocessing steps are available in SI~\ref{SSSNET_appendix:data_description_more}. 

\vspace{-1mm}
\begin{table}[ht]
\vspace{-3mm}
\caption{Summary statistics 
for the real-world networks and one synthetic model. Here $n$ is the number of nodes, $|\mathcal{E}^+|$ and $|\mathcal{E}^-|$ denote the number of positive and negative edges, respectively. $|\Delta^\text{u}|$ counts the number of {unbalanced} triangles (with an odd number of negative edges). 
The {\it violation ratio} 
$\frac{|\Delta^\text{u}|}{|\Delta|}$ (\%) is the percentage of unbalanced triangles in all triangles, i.e., 1 minus the Social Balance Factor from
\cite{patidar2012predicting}.}
\vspace{-2mm}
\centering
\setlength\tabcolsep{2pt}
\begin{tabular}{lrrrrr}
\toprule
Data set               & $n$ &  $|\mathcal{E}^+|$ & $|\mathcal{E}^-|$ & $|\Delta^\text{u}|$ & $\frac{|\Delta^\text{u}|}{|\Delta|}$ (\%) \\ \midrule
Sampson                 & 25&129&126&192&37.16 \\ 
Rainfall&306&64,408&29,228&1,350,756&28.29\\
Fin-YNet &451&14,853&5,431&408,594&26.97\\
S\&P 1500                &1,193&1,069,319&353,930&199,839  &28.15      \\ 
PPI&3,058&7,996&3,864&94 &2.45    \\ 
Syn &5,000&510,586&198,6224&9,798,914&47.20\\
Wiki-Rfa              & 7,634&136,961&38,826&79,911,143     &28.23      \\
\bottomrule
\end{tabular}
\vspace{-5mm}
\label{SSSNET_tab:data_sets}
\end{table}
%%%%%%%%%%%%%%%%%%%%%%%%%%%%%%%%%%%%%%%%%%%%
%%%%%%%%%%%%%%%%%%%%%%%%%%%%%%%%%%%%%%%%%%%%
%%%%%%%%%%%%%%%%%%%%%%%%%%%%%%%%%%%%%%%%%%%%
\vspace{-3mm}
\subsection{Experimental Results}
\label{SSSNET_subsec:results}
In our experiments, we compare {\SSS} against \textbf{nine} state-of-the-art spectral clustering methods in signed networks 
mentioned in Sec.~\ref{SSSNET_sec:related}. 
These methods are based on: 
(1) the symmetric adjacency matrix $\mathbf{A}^*=\frac{1}{2}(\mathbf{A}+(\mathbf{A})^T),$ 
(2) the simple normalized signed Laplacian $\mathbf{\bar{L}}_{sns}=\bar{\mathbf{D}}^{-1}(\mathbf{D}^+-\mathbf{D}^-\mathbf{A}^*)$ and
(3) the balanced normalized signed Laplacian $\mathbf{\bar{L}}_{bns}=\bar{\mathbf{D}}^{-1}(\mathbf{D}^+-\mathbf{A}^*)$ \cite{zheng2015spectral},
(4) the Signed Laplacian matrix $\mathbf{\bar{L}}$ of $\mathbf{A}^*,$
(5) its symmetrically normalized version $\mathbf{L}_\text{sym}$ \cite{kunegis2010spectral}, and the two methods from \cite{Chiang} to optimize the (6) Balanced Normalized Cut and the (7) Balanced Ratio Cut, (8)  SPONGE and 
(9) SPONGE\_sym introduced in \cite{SPONGE_AISTATS_2019},
where the diagonal matrices $\mathbf{\bar{D}}, \mathbf{D}^+$ and $\mathbf{D}^-$ have entries as row-sums of $(\mathbf{A}^*)^++(\mathbf{A}^*)^-, (\mathbf{A}^*)^+$ and $(\mathbf{A}^*)^-$, respectively. 
In our experiments, the abbreviated names of these methods are A, sns, dns, L, L\_sym, BNC, BRC, SPONGE, and SPONGE\_sym, respectively. 
The implementation details are in SI~\ref{SSSNET_appendix:implementation}.

Hyperparameter selection is done via greedy search.
\begin{figure}
    \centering
   \begin{subfigure}[ht]{0.32\linewidth}
  \centering
  \includegraphics[width=\linewidth]{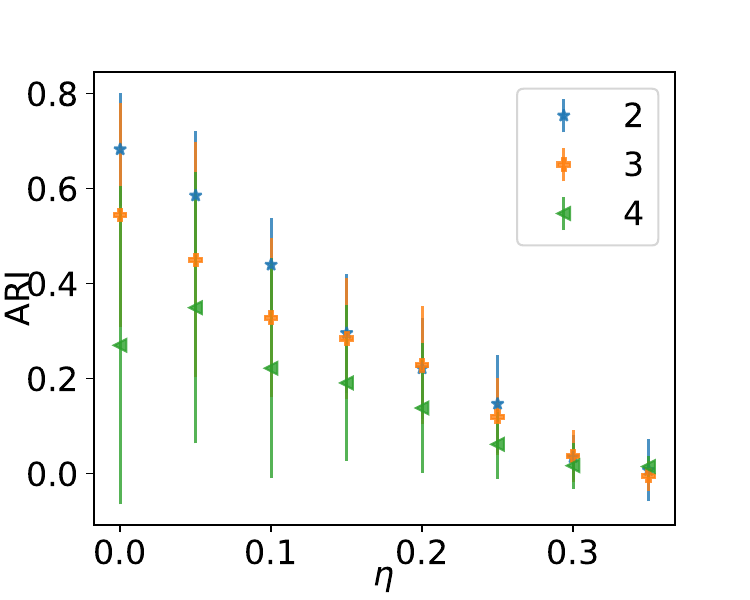}
  \caption{Vary hop $h$ for SIMPA.}
    \end{subfigure}
    \begin{subfigure}[ht]{0.32\linewidth}
      \centering
      \includegraphics[width=\linewidth]{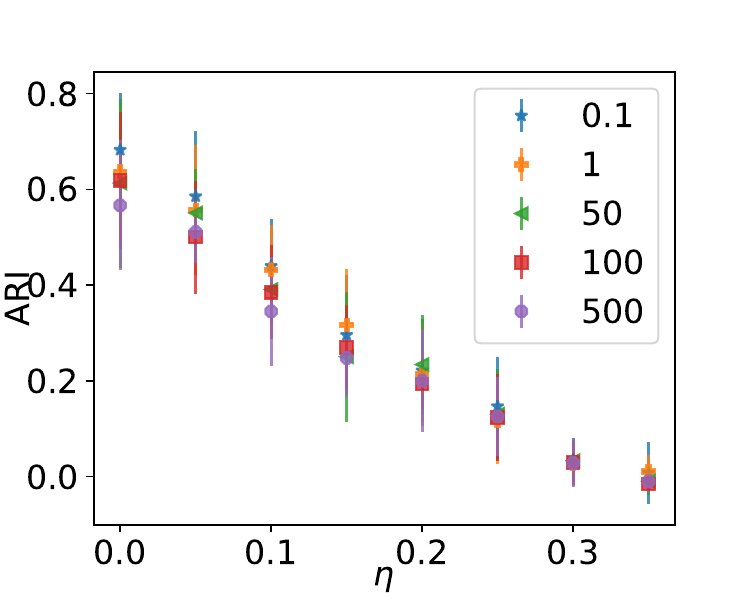}
      \caption{Vary $\gamma_t$ in Eq.~(\ref{SSSNET_eq:loss_overall}).}
      \label{SSSNET_fig:polarized_triplet_loss_ratio}
    \end{subfigure}
        \begin{subfigure}[ht]{0.32\linewidth}
      \centering
      \includegraphics[width=\linewidth]{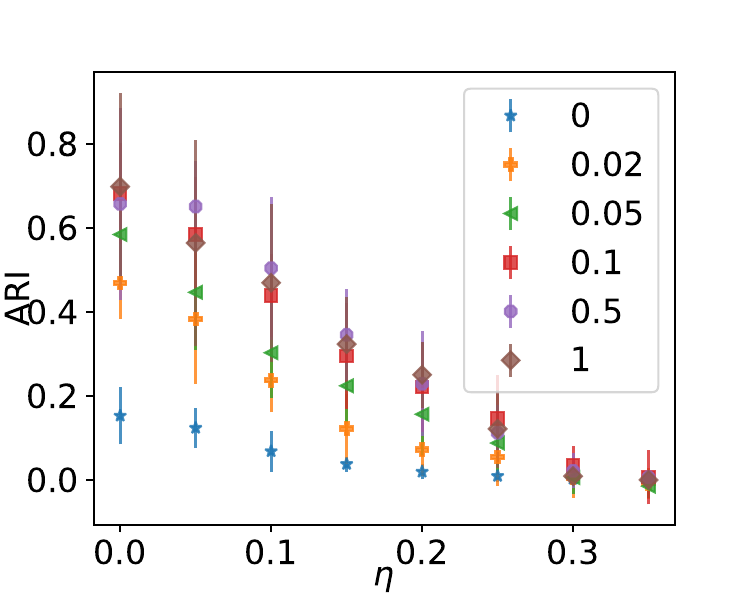}
      \caption{Vary seed ratio.}
      \label{SSSNET_fig:polarized_seed_ratio}
    \end{subfigure}
\begin{subfigure}[ht]{0.32\linewidth}
      \centering
      \includegraphics[width=\linewidth]{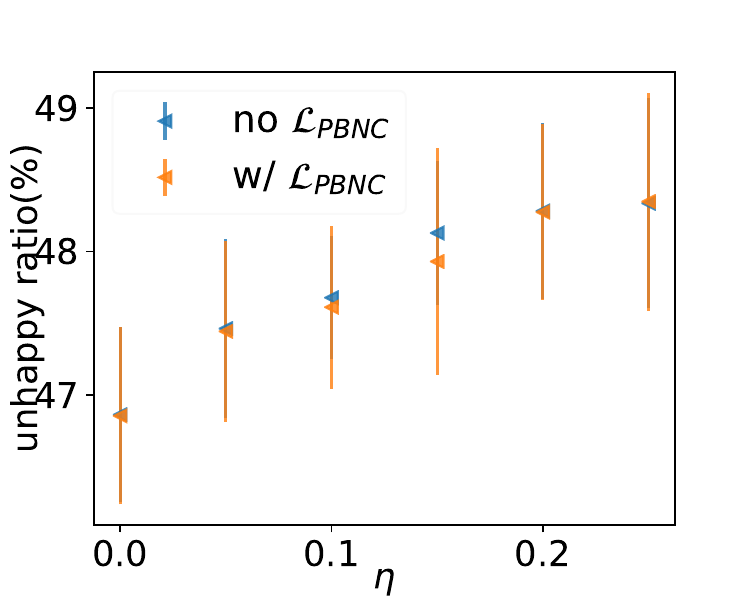}
      \caption{With/without the self-supervised part $\mathcal{L}_\text{PBNC}$ in Eq.~(\ref{SSSNET_eq:loss_overall}).}
      \label{SSSNET_fig:polarized_self_supervised}
    \end{subfigure}
\begin{subfigure}[ht]{0.32\linewidth}
  \centering
      \includegraphics[width=\linewidth]{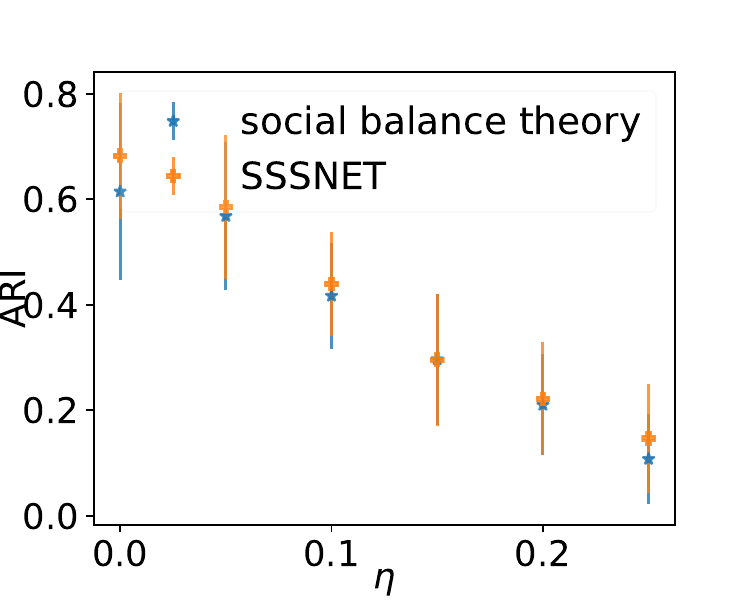}
      \caption{Social balance theory vs {\SSS} (as we vary $\eta$).} 
      \label{SSSNET_fig:polarized_balance}
\end{subfigure}
\begin{subfigure}[ht]{0.32\linewidth}
  \centering
      \includegraphics[width=\linewidth]{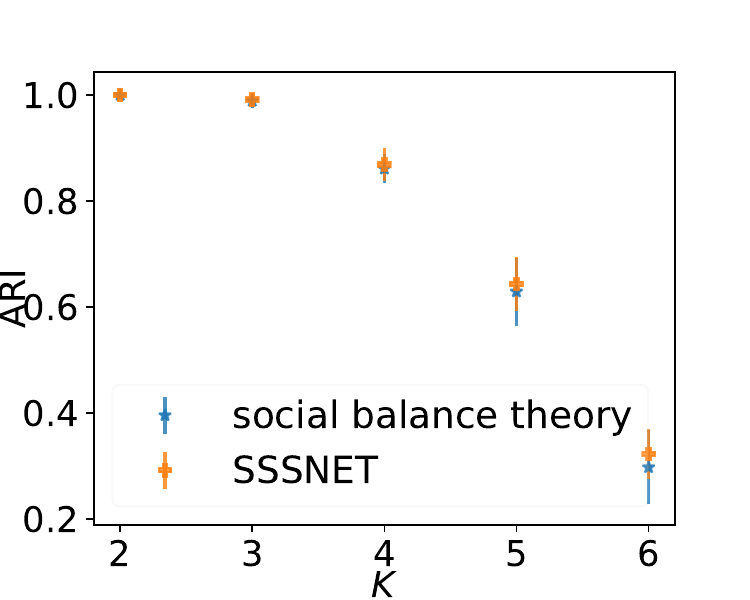}
      \caption{Social balance theory vs {\SSS} (as we vary $K$).}
      \label{SSSNET_fig:SSBM_balance}
\end{subfigure}
    \vspace{-3mm}
    \caption{Hyperparameter analysis (a,b) and ablation study (c-f).
    Figures (a-e) pertain to  \textsc{Pol-SSBM}($n=1050, r =2, p=0.1, \rho=1.5$), 
    while Figure (f) is for an SSBM($n=1000, \eta=0,p=0.01,\rho=1.5$) model with changing $K.$
    Figure (d) compares the ``unhappy ratio" while the others compare the test ARI.
    } 
    \label{SSSNET_fig:ablation}  
\vspace{-8mm}    
\end{figure}
\begin{figure*}[hbt]
    \centering
  \begin{subfigure}[ht]{0.245\linewidth}
    \centering
    \includegraphics[width=\linewidth,trim=0cm 0cm 0cm 1.8cm,clip]{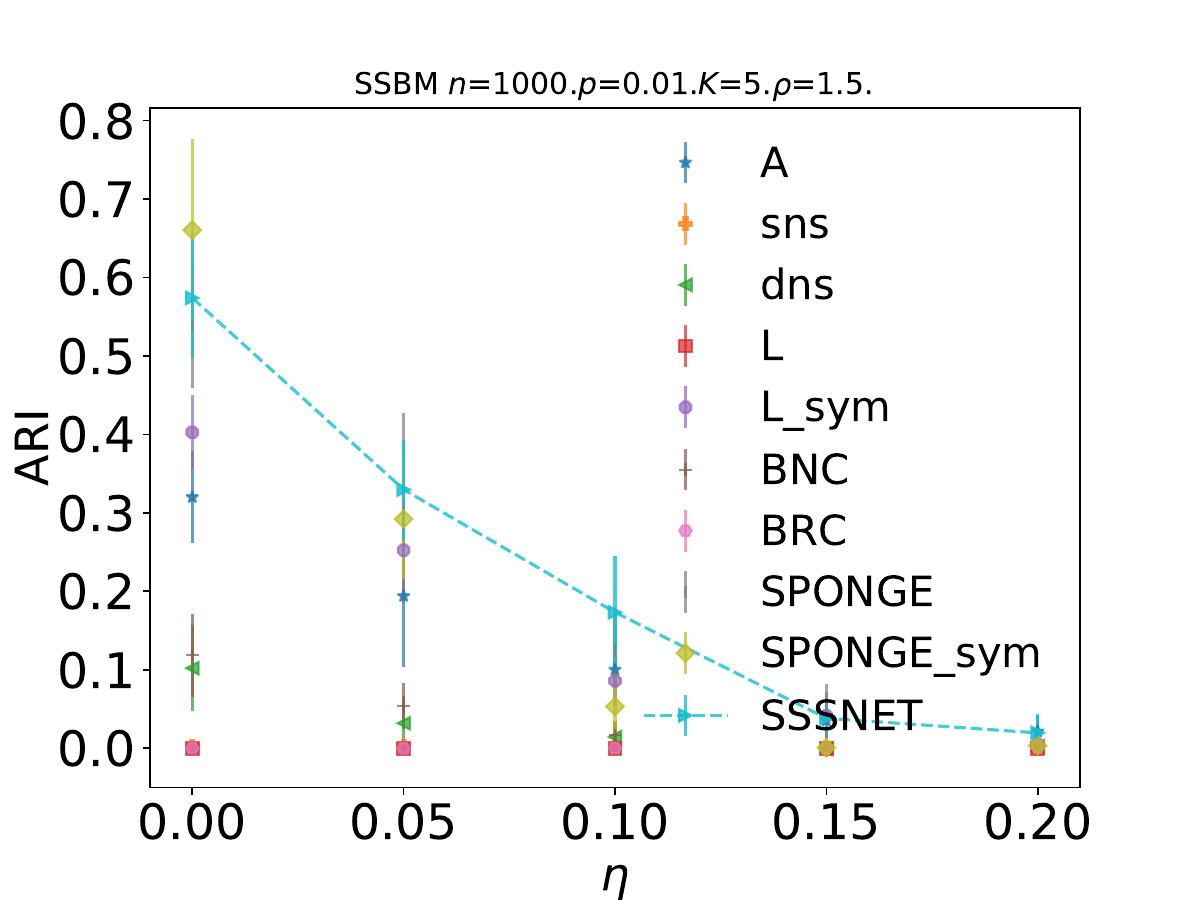}
    \caption{SSBM($n=1000,$\\$ K=5, p=0.01, \rho=1.5$)}
  \end{subfigure}
  \begin{subfigure}[ht]{0.245\linewidth}
    \centering
    \includegraphics[width=\linewidth,trim=0cm 0cm 0cm 1.8cm,clip]{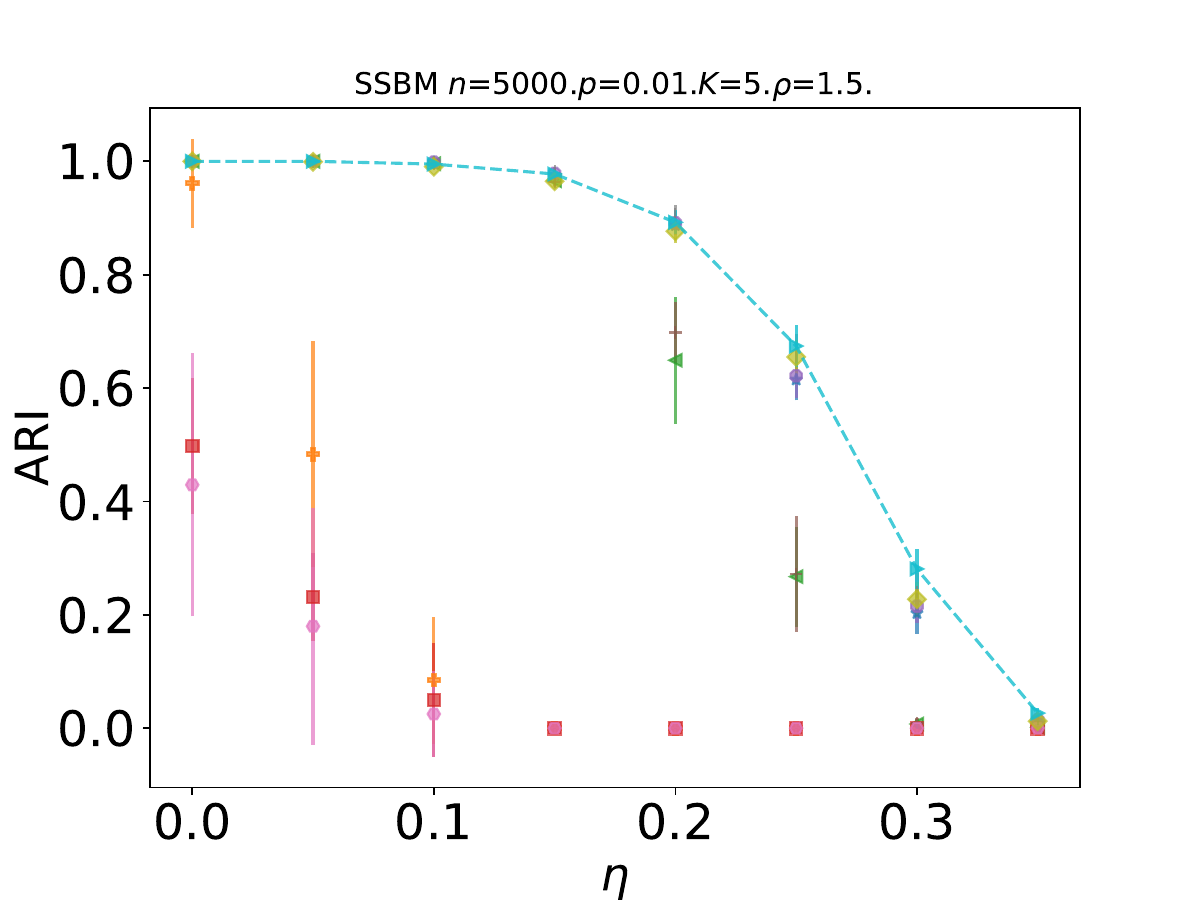}
    \caption{SSBM($n=5000,$\\$ K=5, p=0.01, \rho=1.5$)}
  \end{subfigure}
  \begin{subfigure}[ht]{0.245\linewidth}
    \centering
    \includegraphics[width=\linewidth,trim=0cm 0cm 0cm 1.7cm,clip]{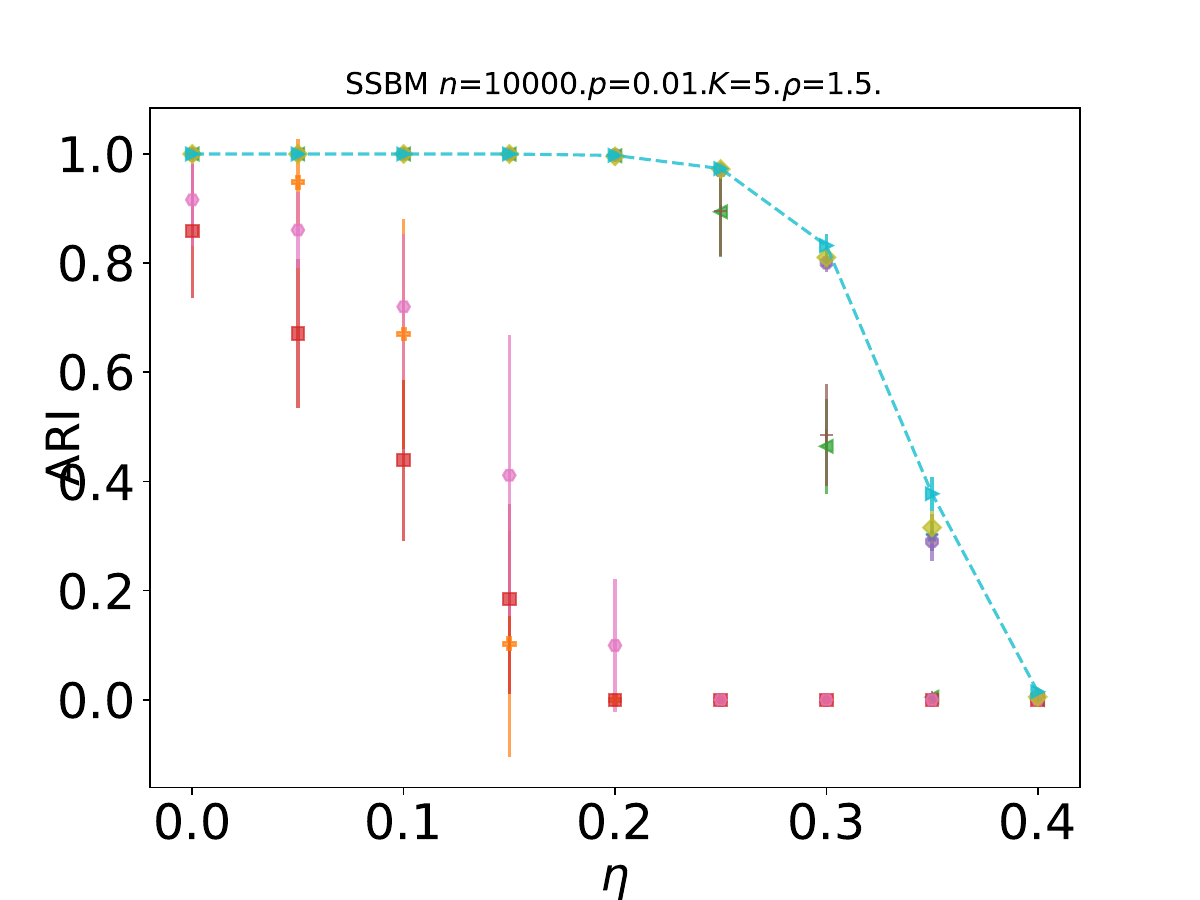}
    \caption{SSBM($n=10000,$\\$ K=5, p=0.01, \rho=1.5)$}
  \end{subfigure}
  \begin{subfigure}[ht]{0.245\linewidth}
    \centering
    \includegraphics[width=\linewidth,trim=0cm 0cm 0cm 1.8cm,clip]{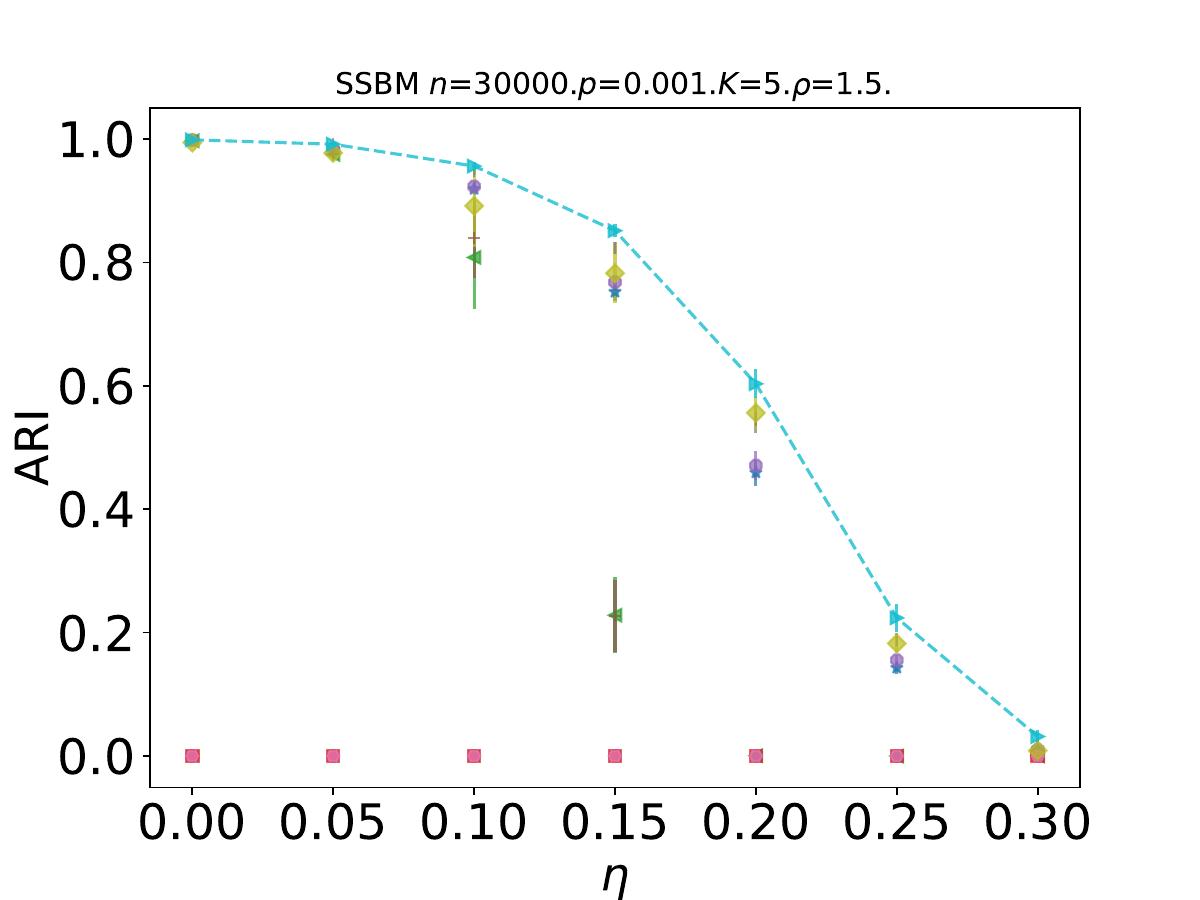}
    \caption{SSBM($n=30000,$\\$ K=5,p=0.001,\rho=1.5$)}
  \end{subfigure}
  \begin{subfigure}[ht]{0.245\linewidth}
    \centering
    \includegraphics[width=\linewidth,trim=0cm 0cm 0cm 1.8cm,clip]{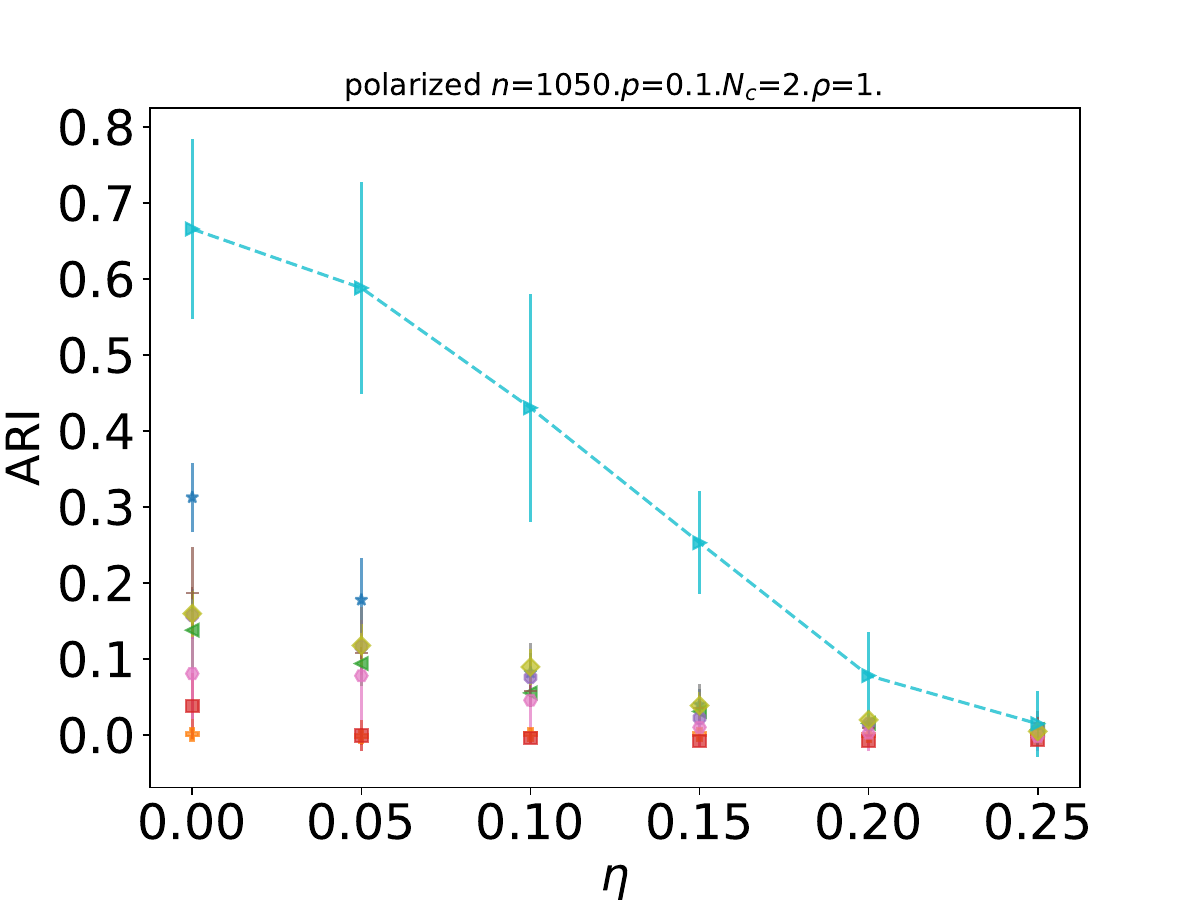}
    \caption{\textsc{Pol-SSBM}($n=1050,$\\$ r=2, p=0.1, \rho=1$) }
  \end{subfigure}
  \begin{subfigure}[ht]{0.245\linewidth}
    \centering
    \includegraphics[width=\linewidth,trim=0cm 0cm 0cm 1.8cm,clip]{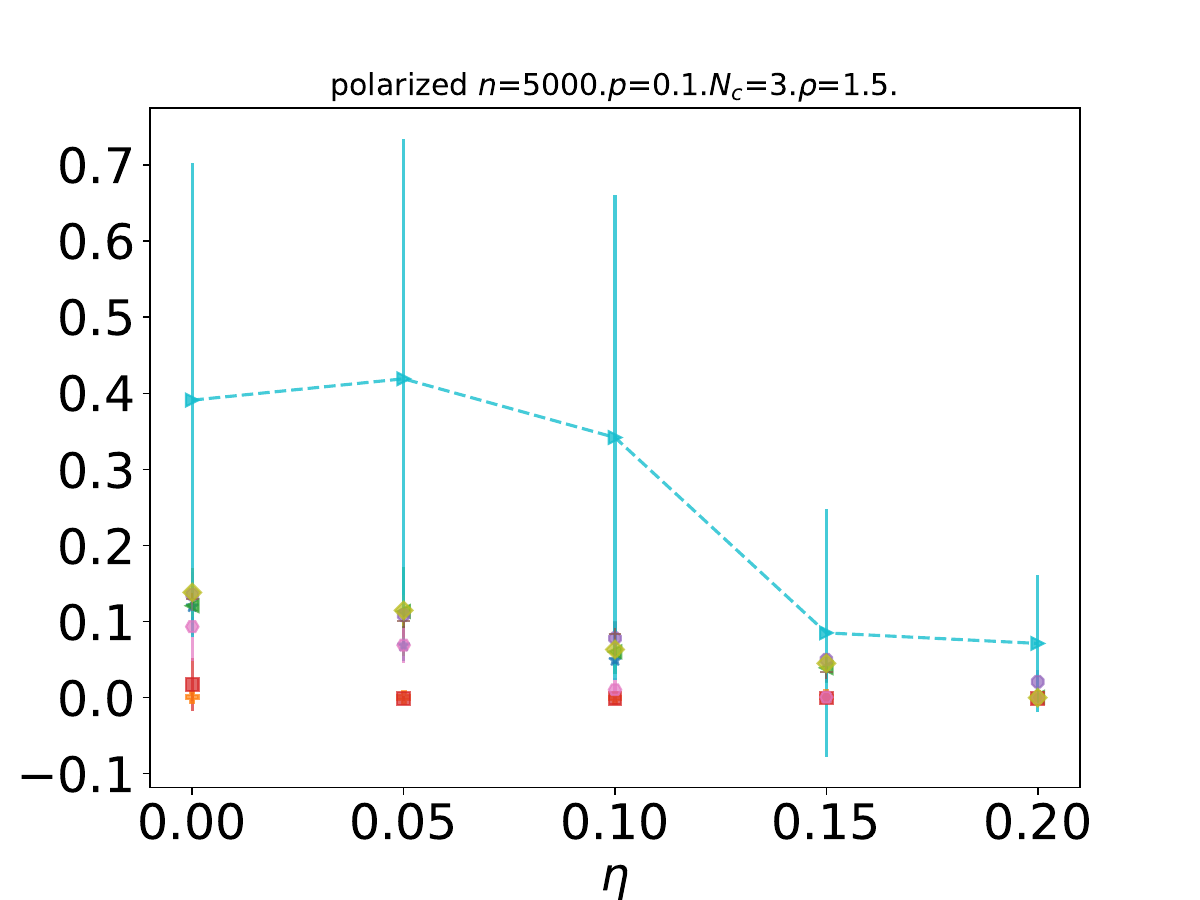}
    \caption{\textsc{Pol-SSBM}($n=5000,$\\$ r=3, p=0.1, \rho=1.5$)}
  \end{subfigure}
  \begin{subfigure}[ht]{0.245\linewidth}
    \centering
    \includegraphics[width=\linewidth,trim=0cm 0cm 0cm 1.8cm,clip]{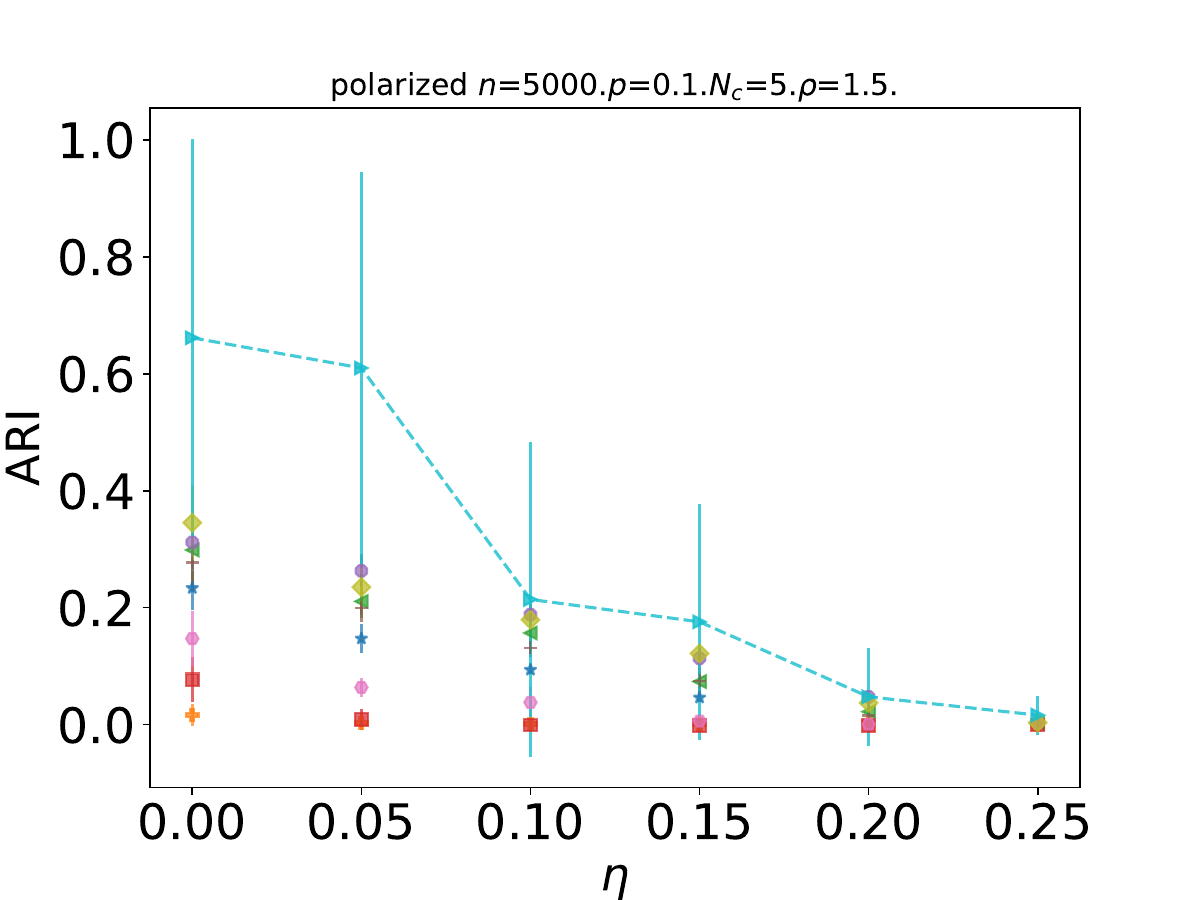}
    \caption{\textsc{Pol-SSBM}($n=5000,$\\$ r=5, p=0.1,\rho=1.5$)}
  \end{subfigure}
  \begin{subfigure}[ht]{0.245\linewidth}
    \centering
    \includegraphics[width=\linewidth,trim=0cm 0cm 0cm 1.8cm,clip]{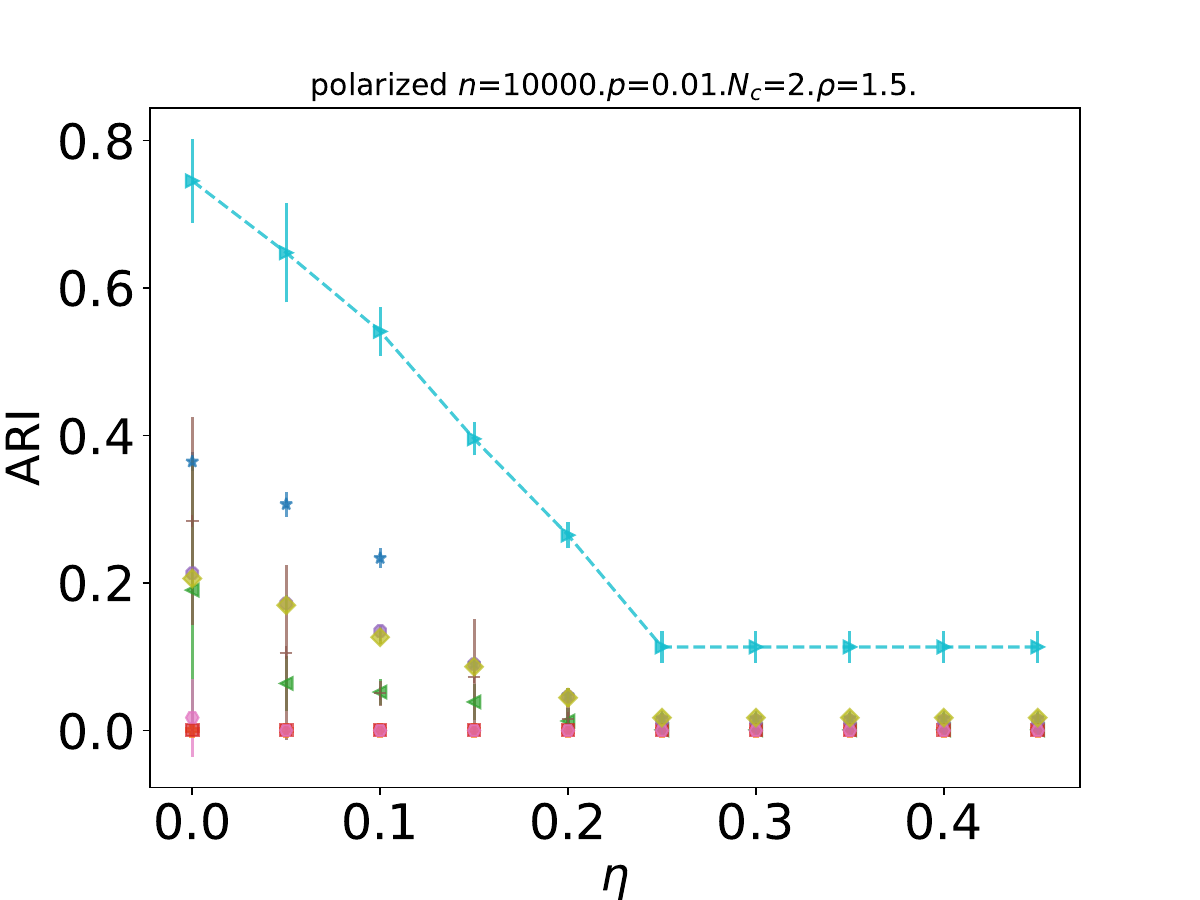}
    \caption{\textsc{Pol-SSBM}($n=10000,$\\$ r  =2, p=0.01, \rho=1.5$)}
  \end{subfigure}
  \vspace{-3mm}
    \caption{Node clustering test ARI comparison on synthetic data. Dashed lines highlight \SSS's performance. 
    Error bars indicate one standard error.
    }
    \label{SSSNET_fig:synthetic_compare}
\end{figure*}
Figure \ref{SSSNET_fig:ablation} (a-b)
compare the performance of {\SSS} on a \textsc{Pol-SSBM}($n=1050, r =2, p=0.1, \rho=1.5$) model under different settings.
We conclude from (a) that as we increase the number of hops to consider, performance drops, which might be explained by too much noise introduced. From (b), we find that the best $\gamma_t$ in our candidates is 0.1. See also SI~\ref{SSSNET_appendix:hyperparameters}.
Our default setting is $h=2, d=32 ,\tau=0.5, \gamma_s=50, \gamma_t = 0.1, \alpha=0.$ 

Unless specified otherwise, we use 10\% of all nodes from each cluster as test nodes, 10\% as validation nodes to select the model, and the remaining 80\% as training nodes, 10\% of which as seed nodes. We train SSSNET for at most 300 epochs with a 100-epoch early-stopping scheme. As \textit{Sampson} is small, 50\% nodes are training nodes, 50\% of which are seed nodes, and no validation nodes; 
we train 80 epochs and test on the remaining 50\% nodes. For \textit{S\&P 1500}, \textit{Fin-YNet}  and \textit{Rainfall}, we use 90\% nodes for training, 10\% of which as seed nodes, and no validation nodes for 300 epochs. 
If node attributes are missing, {\SSS} stacks the eigenvectors corresponding to the largest $K$ eigenvalues of the symmetrized adjacency matrix $\frac{1}{2}(\mathbf{A}+\mathbf{A}^T)$ as $\mathbf{X}_\mathcal{V}$ for synthetic data, and 
the eigenvectors corresponding to the smallest $K$ eigenvalues of the symmetrically normalized Signed Laplacian~\cite{hou2003laplacian} 
of $\frac{1}{2}(\mathbf{A}+\mathbf{A}^T)$ for real-world data. 
Numerical results are averaged over 10 runs; 
error bars indicate one standard error.

\vspace{-5mm}
\subsubsection{Node Clustering Results on Synthetic Data}

Figure \ref{SSSNET_fig:synthetic_compare} compares the numerical performance of {\SSS} with other methods on synthetic data. 
We remark that {\SSS} gives state-of-the-art test ARIs on a wide range of network densities and noise levels, on various network scales, 
especially for polarized SSBMs. SI~\ref{SSSNET_appendix_subsec:synthetic_more} 
provides more results with different measures.

\vspace{-2mm}
%%%%%%%%%%%%%%%%%%%%%%%%%%%%%%%%%%%%%%%%%%%%
\subsubsection{Node Clustering Results on Real-World Data}
\begin{table*}
\vspace{-1mm}
\small
  \caption{Clustering performance  
  on real-world data sets; best is 
  in \red{bold red}, and $2^\text{nd}$ in \blue{underline blue}. The first 3 rows are test ARIs, the $4^\text{th}$ ``ARI distance to best", the rest are unhappy ratios (\%).
  }
  \vspace{-2mm} 
  \label{SSSNET_tab:real_data_performance}
  \setlength\tabcolsep{0.2pt}
  \begin{tabular}{c c c c  c  cccccc}
    \toprule
    Data set&A&sns&dns&L&L\_{sym}&BNC&BRC&SPONGE&SPONGE\_{sym}&SSSNET\\
    \midrule
    %\multirow{3}{*}{ARI}&
    Sampson& 0.32$\pm$0.10&0.15$\pm$0.09&0.33$\pm$0.10  &0.16$\pm$0.05&0.35$\pm$0.09&0.32$\pm$0.12&0.21$\pm$0.11&\blue{0.36$\pm$0.11}&0.34$\pm$0.11&\red{0.55$\pm$0.07}\\
    Rainfall&0.61$\pm$0.08&0.28$\pm$0.03&0.65$\pm$0.04&0.46$\pm$0.06&0.58$\pm$0.07&0.62$\pm$0.05&0.47$\pm$0.05&N/A&\blue{0.75$\pm$0.09}&\red{0.76$\pm$0.13}\\
    S\&P 1500&0.21$\pm$0.00&0.00$\pm$0.00&0.05$\pm$0.01&0.06$\pm$0.00&0.24$\pm$0.00&0.04$\pm$0.00&0.00$\pm$0.00&0.30$\pm$0.00&\blue{0.34$\pm$0.00}&\red{0.66$\pm$0.00}\\
    \midrule
    Fin-YNet & 0.22$\pm$0.09 & 0.37$\pm$0.12 & 0.32$\pm$0.10
    & 0.33$\pm$0.10 & 0.22$\pm$0.09 & 0.32$\pm$0.09 & 0.33$\pm$0.11 & 0.20$\pm$0.08 & \blue{0.16$\pm$0.07} & \red{0.00$\pm$0.00} \\
    \midrule
    PPI&  57.59$\pm$0.55&46.82$\pm$0.01&46.79$\pm$0.04 & 46.91$\pm$0.03 & 47.05$\pm$0.04 & 46.63$\pm$0.04 & 52.11$\pm$0.42 & 47.57$\pm$0.00 & \blue{46.39$\pm$0.10} & \red{17.64$\pm$0.84}\\
   Wiki-Rfa&   50.05$\pm$0.03&23.28$\pm$0.00&23.28$\pm$0.00 & 23.28$\pm$0.00 & 36.95$\pm$0.01 & 23.28$\pm$0.00 & 23.49$\pm$0.00 & 29.63$\pm$0.01 & \red{23.26$\pm$0.00} & \blue{23.27$\pm$0.14}\\
  \bottomrule
\end{tabular}
\end{table*}
\vspace{-1mm}
As Table \ref{SSSNET_tab:real_data_performance} shows, {\SSS} yields the most accurate cluster assignments on \textit{S\&P 1500}, and \textit{Sampson} 
in terms of test ARI, compared to baselines.
When using labels from SPONGE to conduct semi-supervised training, {\SSS} also achieves the most accurate test ARI. The N/A entry for SPONGE denotes that we use it as ``ground-truth", so do not compare {\SSS} against SPONGE on \textit{Rainfall}. 
``ARI dist. to best" on \textit{Fin-YNet} considers the average distance of test ARI performance to the best average test ARI for each of the 21 years, and then obtains mean and standard deviation over the 21 distances for each method. We conclude that {\SSS} produces the highest test ARI for all 21 years. 
For data sets without labels, we compare {\SSS} with other methods in terms of ``unhappy ratio", the ratio of unhappy edges. 
We conclude that {\SSS} gives comparable and often better results on these data sets, in a self-supervised setting. SI~\ref{SSSNET_appendix_subsec:real_more} provides extended results.

\vspace{-3mm}
\subsubsection{Ablation Study}
\label{SSSNET_sec:AblationStudy}
In Figure \ref{SSSNET_fig:ablation}, (c-e) 
rely on  \textsc{Pol-SSBM}($n=1050, r =2, p=0.1, \rho=1.5$), while (f) is based on an SSBM($n=1000, \eta=0,p=0.01,\rho=1.5$) model with varying  $K$.  (c) explores the influence of increasing seed ratio, and validates our strength in using labels. (d) assesses the impact of removing the self-supervised loss $\mathcal{L}_\text{PBNC}$ from Eq.~(\ref{SSSNET_eq:loss_overall}).
Lower values of the ``unhappy ratio" with $\mathcal{L}_\text{PBNC}$ reveal that including the self-supervised loss in Eq.~(\ref{SSSNET_eq:loss_overall}) can be beneficial.

Figure \ref{SSSNET_fig:ablation} (e)
compares the performance for $h=2$ by replacing SIMPA with an aggregation scheme based on social balance theory, which also considers a path of length two with pure negative links as a path of friendship. The ARI for SSSNET is larger than the one for the corresponding method which would adhere to social balance theory, thus further validating our approach.
Figure \ref{SSSNET_fig:ablation} (f) further illustrates the influence of the violation ratio, as percentage (the last column in Table \ref{SSSNET_tab:data_sets}) on the performance gap between a variant based on social balance theory and our model.
As $K$ increases, the violation ratio
grows, and we witness a larger gap in test ARI performance, suggesting  that social balance theory becomes more problematic when there are more violations to its assumption that ``an enemy's enemy is a friend".
We conclude that this social balance theory modification not only complicates the calculation by adding one more scenario of friendship, but can also decrease the test ARI.
Finally, as we increase the ratio of seed nodes, we witness an increase in test ARI performance, as expected.

\vspace{-3mm}
%%%%%%%%%%%%%%%%%%%%%%%%%%%%%%%%%%%%%%
\section{Conclusion and Future Work}
\label{SSSNET_sec:conclusion}
{\SSS} 
provides an end-to-end pipeline to create node embeddings and carry out signed clustering, with or without available additional node features, and with an emphasis on polarization. 
It would be interesting to apply the method to more networks without ground truth
and relate the resulting clusters to exogenous information. 
As another future direction, 
we aim
to extend our framework to also detect the number of clusters, see e.g., 
\cite{riolo2017efficient}. 
Other future research directions will address 
the performance in the very sparse regime, where spectral methods underperform  and various regularization techniques have been proven to be effective both on the theoretical and experimental fronts; for example, see the regularization in the sparse regime for the unsigned \cite{chaudhuri12,Amini_2013} and signed clustering  settings \cite{SPONGE2020regularized}. 
Applying signed clustering to cluster multivariate time series, and leveraging the uncovered clusters for the time series prediction task, by fitting the model of choice for each individual cluster, as in  \cite{jha2015clusteringToForecast}, is a promising extension. Finally, 
adapting our pipeline for constrained clustering, a popular task in the semi-supervised learning \cite{consClust}, is worth exploring.
%%%%%%%%%%%%%%%%%%%%%%%%%%%%%%%%%%%%%%%
\paragraph{Acknowledgements.} 
YH is supported by a Clarendon scholarship. 
GR is funded in part by EPSRC grants EP/T018445/1 and EP/R018472/1. MC acknowledges support from the EPSRC grant EP/N510129/1 at The Alan Turing Institute.
%%%%%%%%%%%%%%%%%%%%%%%%%%%%%%%%%%%%%%%%%
% references
%%%%%%%%%%%%%%%%%%%%%%%%%%%%%%%%%%%%%%%%
\vspace{-2mm}
\bibliographystyle{plain}
{\small
\bibliography{SDM_arxiv}}
\appendix

%%%%%%%%%%%%%%%%%%%%%%%%%%%%%%%%%%%%%%%%%%%%%%%%%%%%
% more results
%%%%%%%%%%%%%%%%%%%%%%%%%%%%%%%%%%%%%%%%%%%%%%%%%%%%%%
\section{Additional Results}
\label{SSSNET_appendix:additional_results}
\subsection{Additional Results on Synthetic Data}
\label{SSSNET_appendix_subsec:synthetic_more}
\begin{figure*}[ht]
    \centering
    \begin{subfigure}[ht]{0.24\linewidth}
    \centering
    \includegraphics[width=\linewidth,trim=0cm 0cm 0cm 1.7cm,clip]{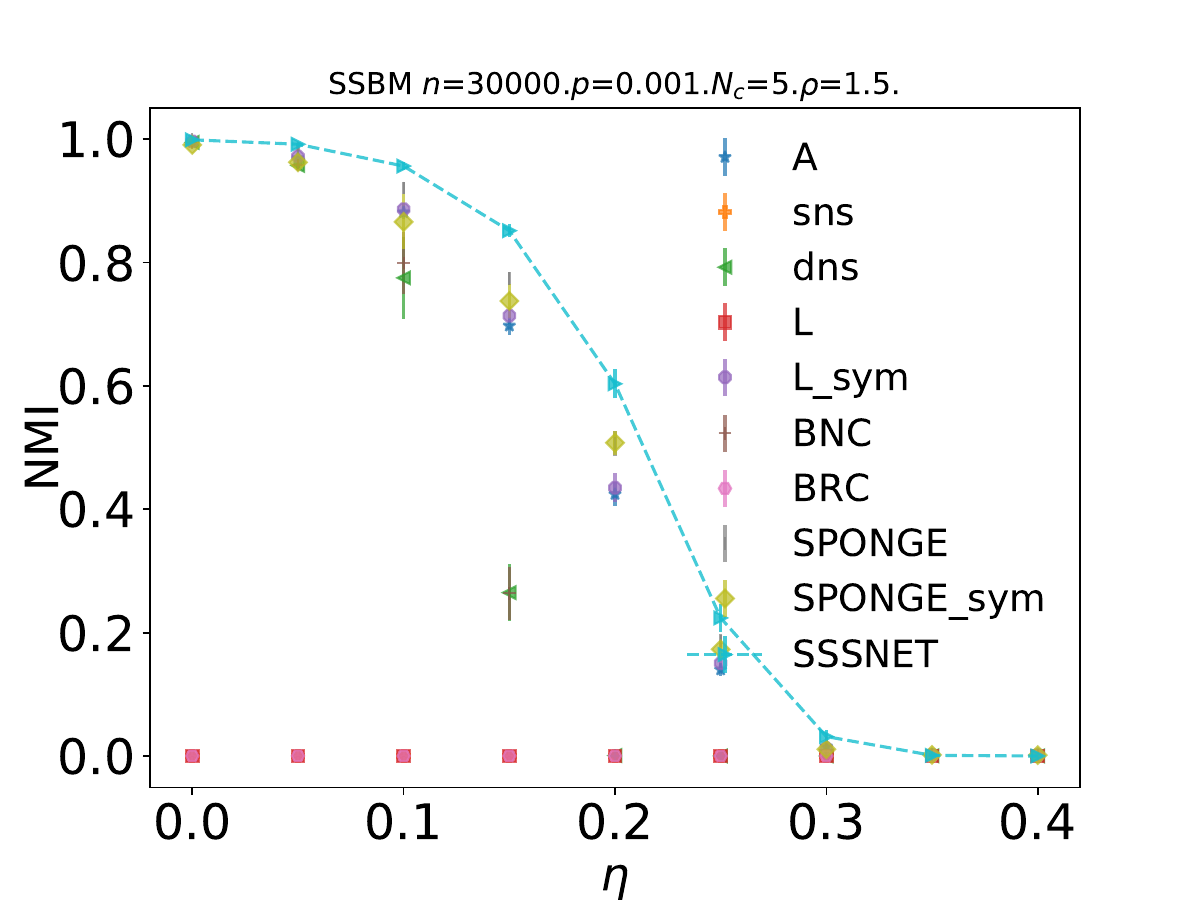}
    \captionsetup{width=1.0\linewidth}
    \caption{NMI of SSBM($n=30000, K=5, p=0.001, \rho=1.5$)}
  \end{subfigure}
  \begin{subfigure}[ht]{0.24\linewidth}
    \centering
    \includegraphics[width=\linewidth,trim=0cm 0cm 0cm 1.7cm,clip]{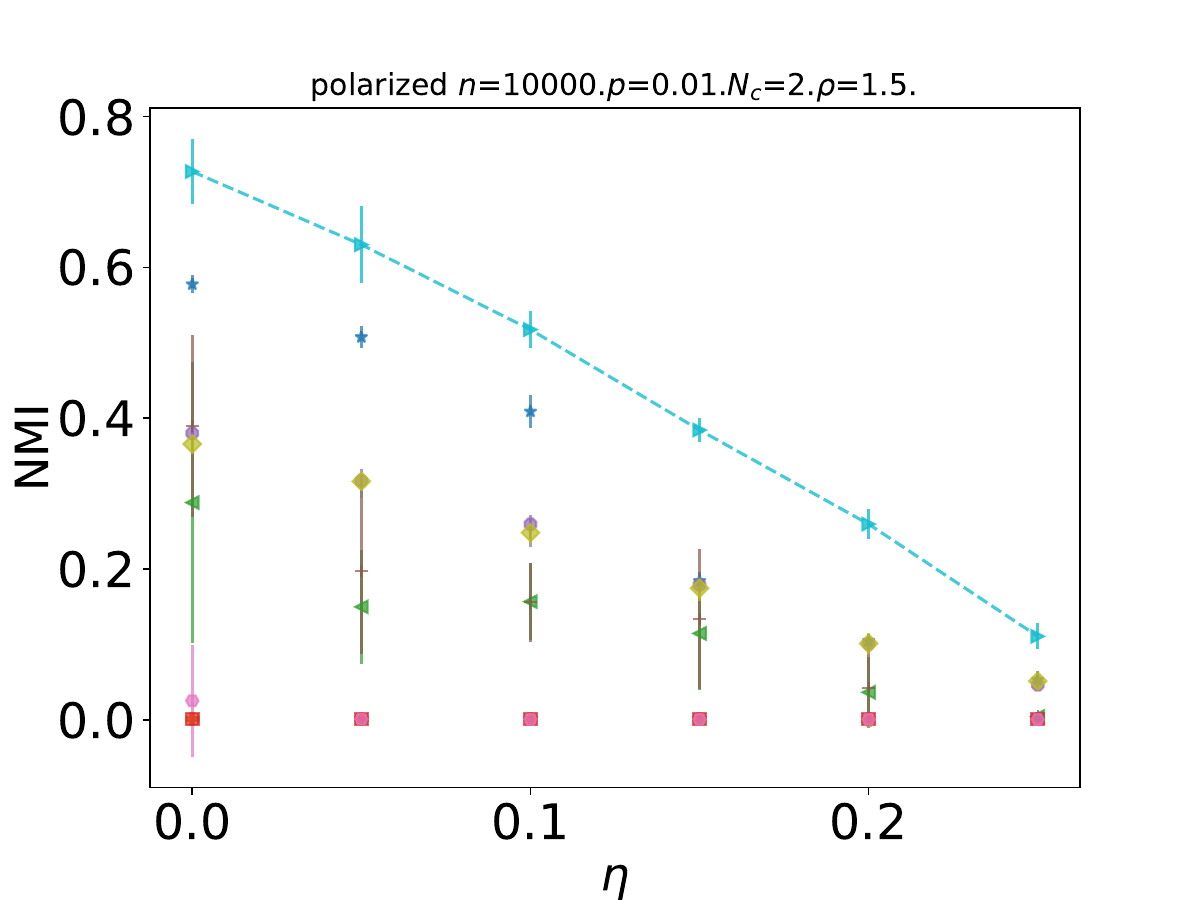}
    \captionsetup{width=1.0\linewidth}
    \caption{NMI of \textsc{Pol-SSBM}($n=10000, n_{c}=2, p=0.01, \rho=1.5$)}
  \end{subfigure}
  \begin{subfigure}[ht]{0.24\linewidth}
    \centering
    \includegraphics[width=\linewidth,trim=0cm 0cm 0cm 1.7cm,clip]{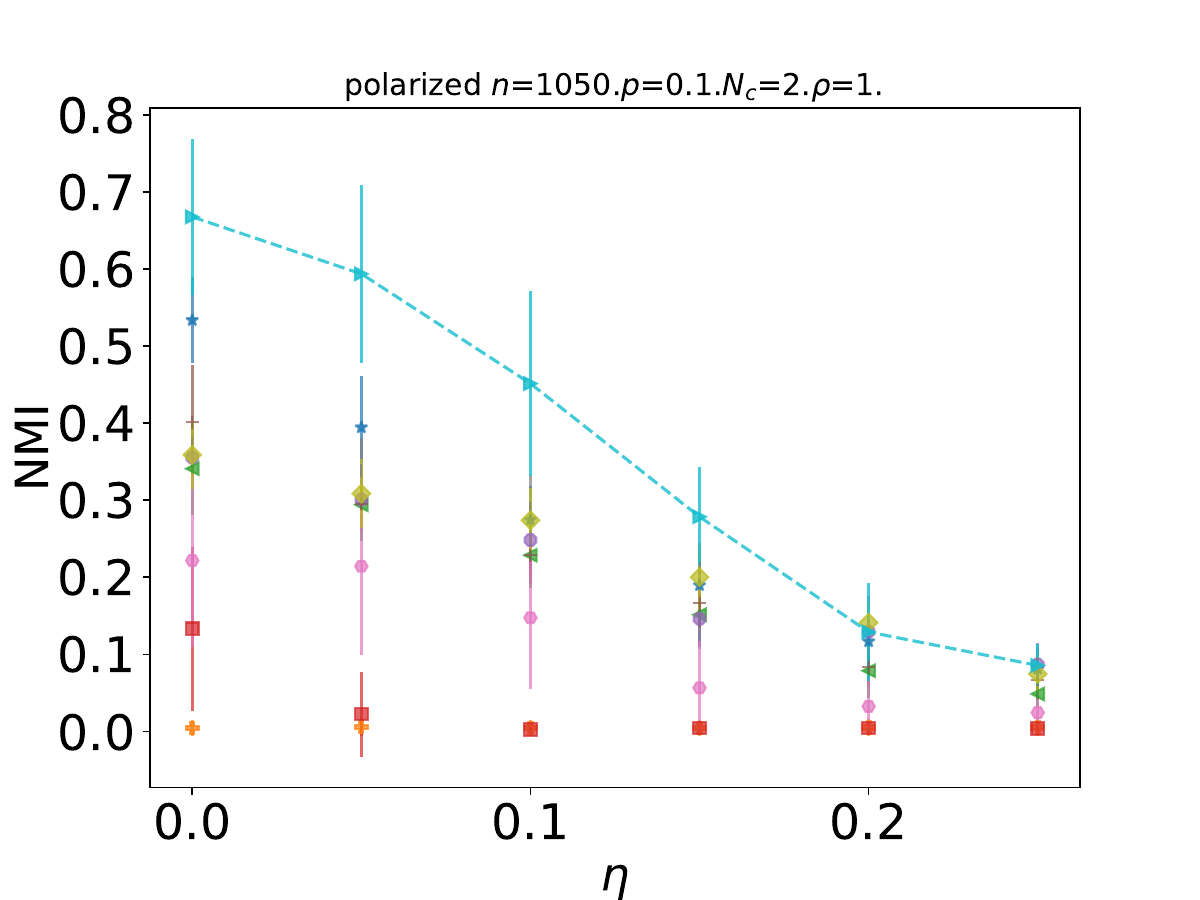}
    \captionsetup{width=1.0\linewidth}
    \caption{NMI of\\ \textsc{Pol-SSBM}($n=1050, n_{c}=2, p=0.1, \rho=1$)}
  \end{subfigure}
\begin{subfigure}[ht]{0.24\linewidth}
    \centering
    \includegraphics[width=\linewidth,trim=0cm 0cm 0cm 1.7cm,clip]{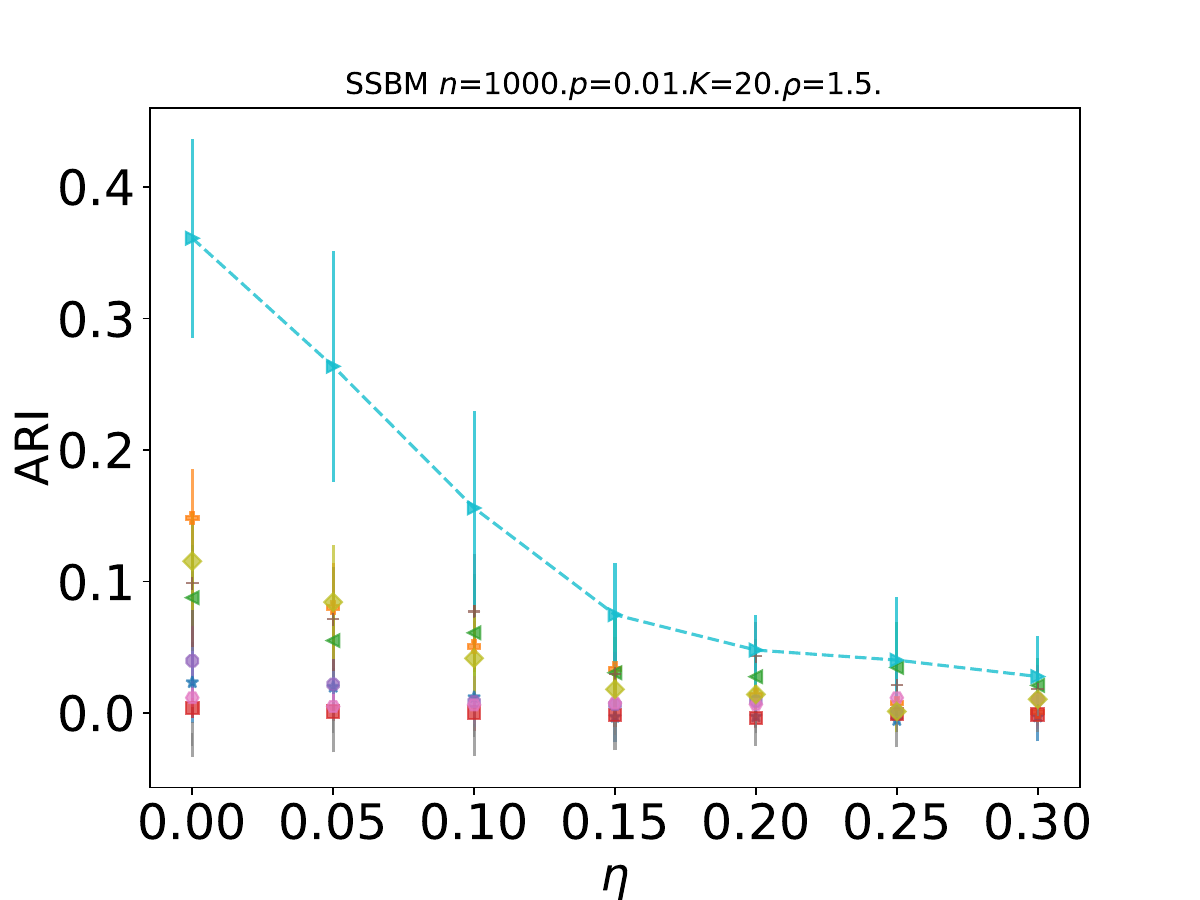}
    \captionsetup{width=1.0\linewidth}
    \caption{ARI of SSBM($n=1000, K=20, p=0.01, \rho=1.5$)}
  \end{subfigure}
  \begin{subfigure}[ht]{0.24\linewidth}
    \centering
    \includegraphics[width=\linewidth,trim=0cm 0cm 0cm 1.7cm,clip]{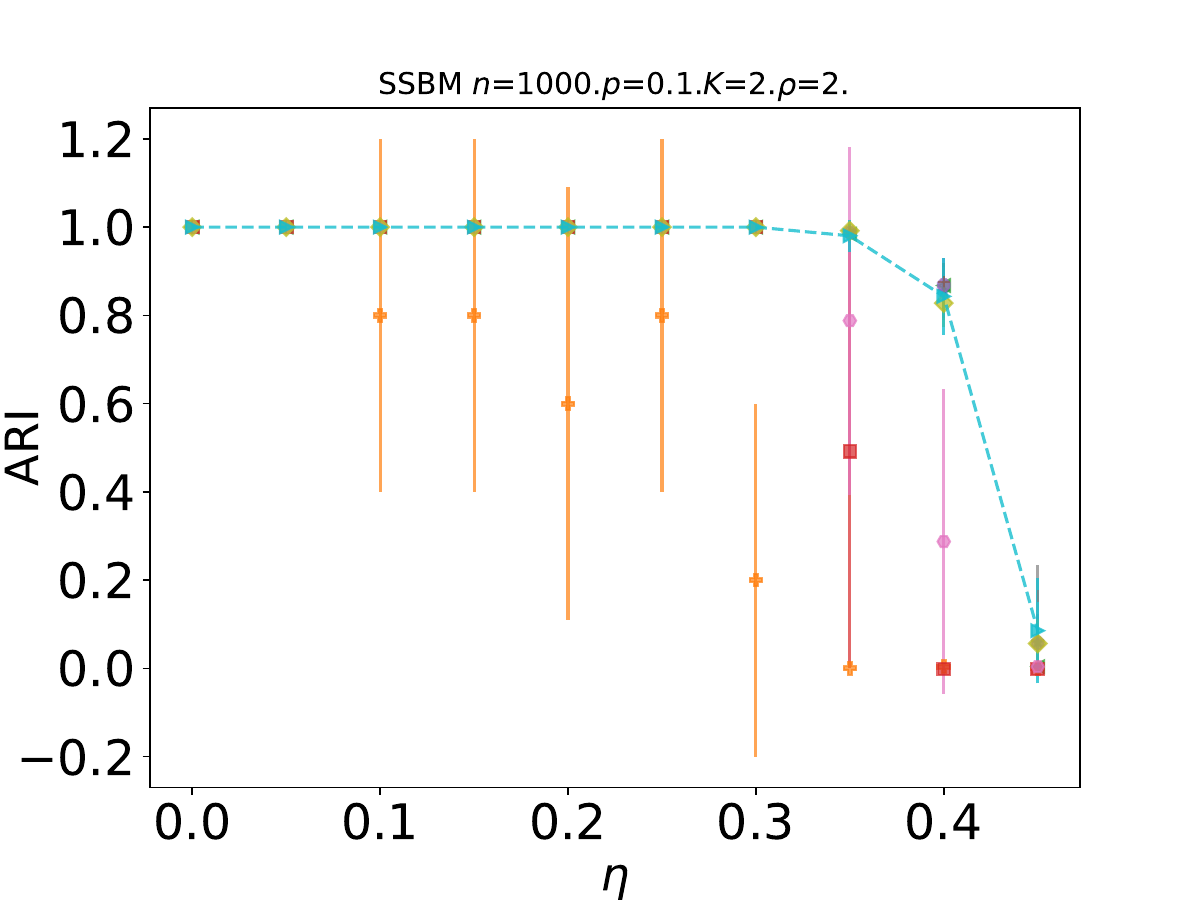}
    \captionsetup{width=1.0\linewidth}
    \caption{ARI of \\
    SSBM($n=1000, K=2, p=0.1, \rho=2$)}
  \end{subfigure}
    \begin{subfigure}[ht]{0.24\linewidth}
    \centering
    \includegraphics[width=\linewidth,trim=0cm 0cm 0cm 1.7cm,clip]{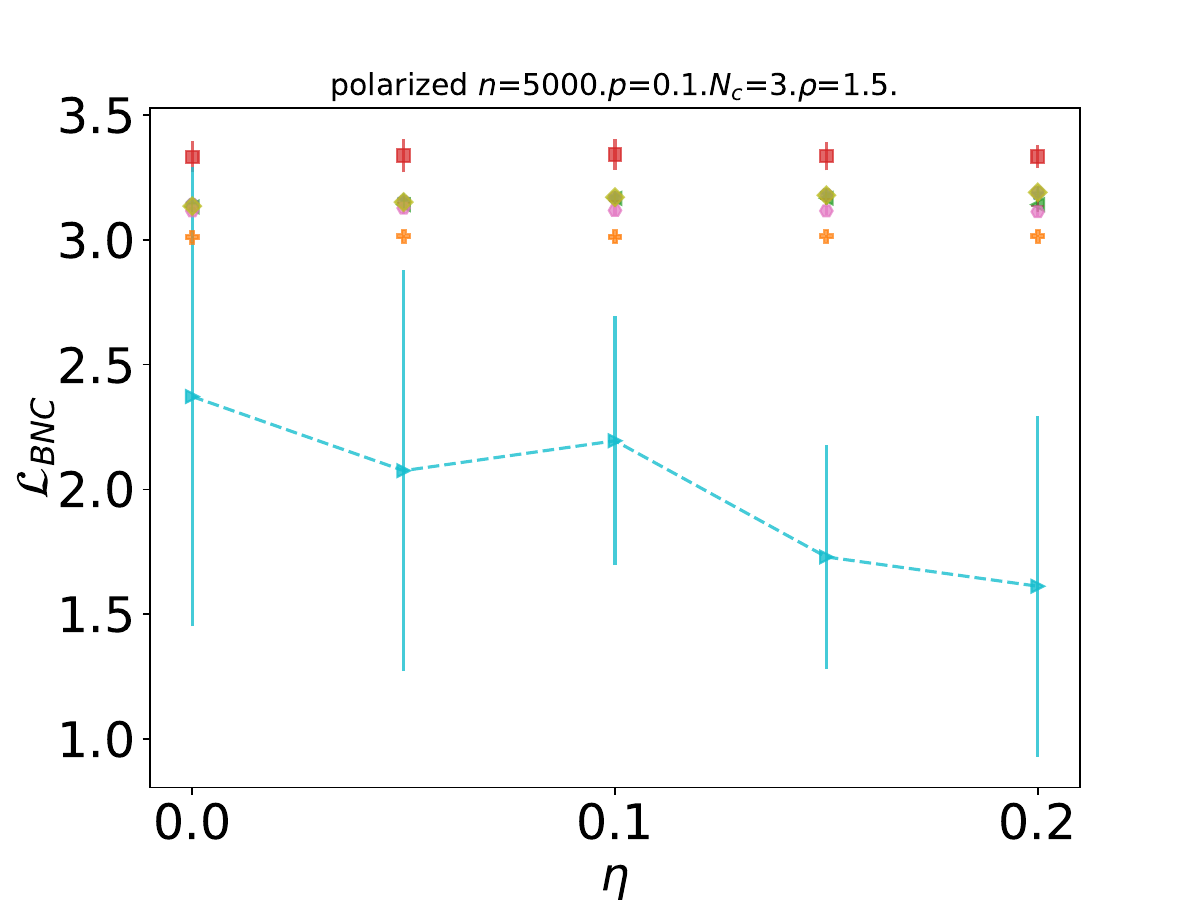}
    \captionsetup{width=1.0\linewidth}
    \caption{$\mathcal{L}_\text{BNC}$ of \textsc{Pol-SSBM}($n=5000, n_{c}=3, p=0.1, \rho=1.5$)}
  \end{subfigure}
  \begin{subfigure}[ht]{0.24\linewidth}
    \centering
    \includegraphics[width=\linewidth,trim=0cm 0cm 0cm 1.7cm,clip]{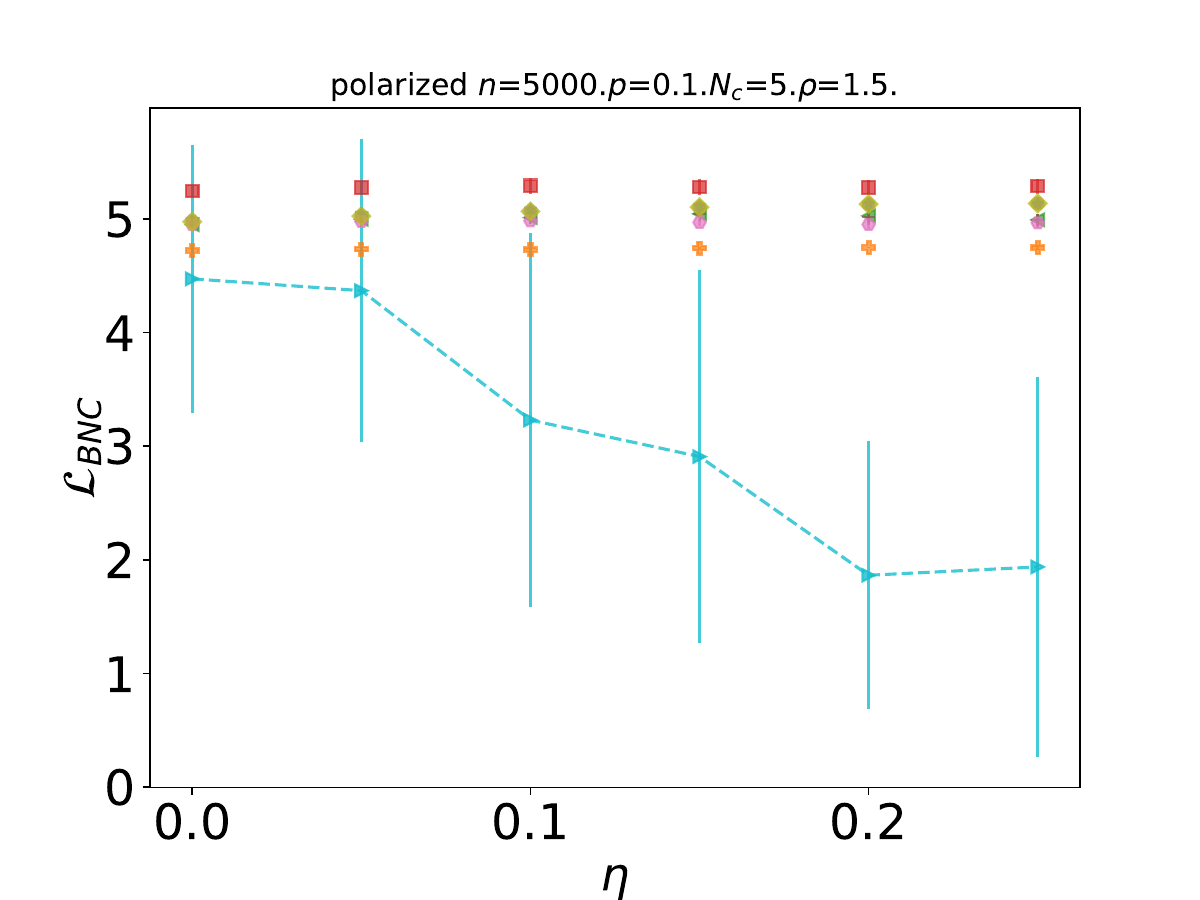}
    \captionsetup{width=1.0\linewidth}
    \caption{$\mathcal{L}_\text{BNC}$ of \textsc{Pol-SSBM}($n=5000, n_{c}=5, p=0.1, \rho=1.5$)}
  \end{subfigure}
  \begin{subfigure}[ht]{0.24\linewidth}
    \centering
    \includegraphics[width=\linewidth,trim=0cm 0cm 0cm 1.7cm,clip]{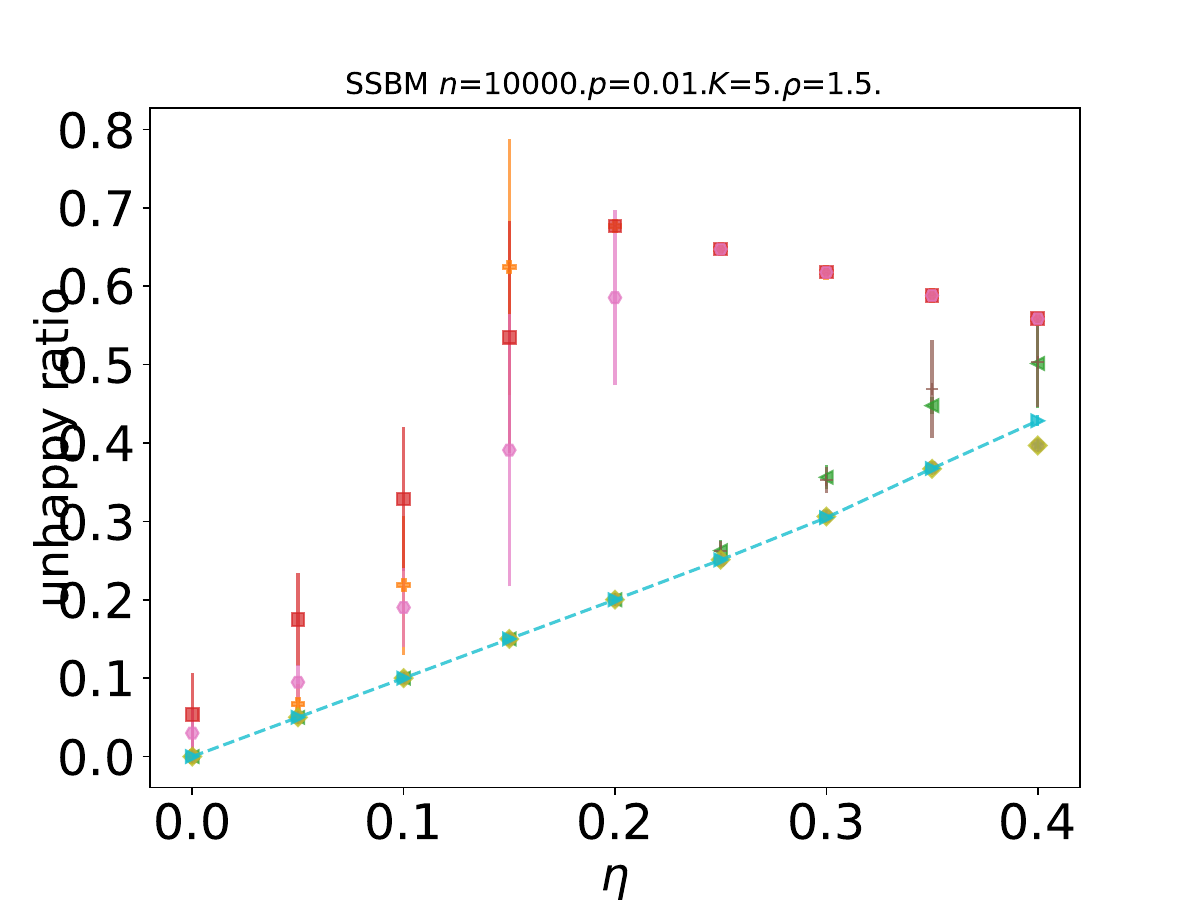}
    \captionsetup{width=1.0\linewidth}
    \caption{Unhappy ratio of SSBM($n=10000, K=5, p=0.01, \rho=1.5$)}
  \end{subfigure}
    \caption{Extended node clustering result comparison on synthetic data. Dashed lines are added to {\SSS}'s performance to highlight our result. Each setting is averaged over ten runs. Error bars are given by standard errors. NMI and ARI results are on test nodes only (the higher, the better), while $\mathcal{L}_\text{BNC}$ and unhappy ratio results are on all nodes in the signed network (the lower, the better).
    }
    \label{SSSNET_fig:synthetic_compare_more}
\end{figure*}

Figure \ref{SSSNET_fig:synthetic_compare_more} provides further  results on synthetic data.
In addition to the ARI scores for two more synthetic settings, SSBM($n=1000, K=20, p=0.01, \rho=1.5$) and SSBM($n=1000, K=2, p=0.1, \rho=2$), we also report the NMI scores, Balanced Normalized Cut values $\mathcal{L}_\text{BNC},$ and unhappy ratios, on some synthetic data used in Figure~\ref{SSSNET_fig:synthetic_compare} 
in the main text.
From Figure \ref{SSSNET_fig:synthetic_compare_more}, we remark that {\SSS} gives comparable balanced normalized cut values and unhappy ratios in these regimes, and leading performance in terms of both ARI and NMI.
%%%%%%%%%%%%%%%%%%%%%%%%%%%%%%%%%%%%%%%%%%%%%%%%
\subsection{Additional Results on Real-World Data}
\label{SSSNET_appendix_subsec:real_more}
%%%%%%%%%%%%%%%%%%%%%%%%%%%%%%%%%%%
\vspace{-3mm}
\subsubsection{Discussion on Attributes for \textit{Sampson}}
On this data set, {\SSS} with the `Cloisterville' attribute achieves highest ARI. When ignoring this attribute and instead using the identity matrix with 25 rows as input feature matrix for \textit{Sampson}, we achieve a test ARI 0.37$\pm$0.19, which is much lower than {\SSS}'s test ARI with 1-dimensional attributes, but still higher than the other methods.
%%%%%%%%%%%%%%%%%%%%%%%%%%%%%%%%%%%%%%%
\vspace{-3mm}
\subsubsection{Extended Result Table for \textit{Fin-YNet}}
Table \ref{SSSNET_tab:real_data_performance_full} gives an extended comparison of different methods on the financial correlation data set \textit{Fin-YNet}. In the first panel of table the difference ($\pm$ 1 s.e. when applicable) to the best-performing method is given; hence each row will have at least one zero entry. 
\begin{table*}
\vspace{-1mm}
  \caption{Clustering performance comparison on \textit{Fin-YNet}; the first panel shows distance to the best performance and the second panel shows absolute performance. The best is in {\red{bold red}}, and second best in {\blue{underline blue}}.
  Standard deviations are not shown due to space constraint.
  }
  \vspace{-2mm} 
  \label{SSSNET_tab:real_data_performance_full}
  \centering
  \begin{tabular}{c c c c  c  cccccc}
    \toprule
    Metric&A&sns&dns&L&L\_{sym}&BNC&BRC&SPONGE&SPONGE\_{sym}&SSSNET\\
    \midrule
    test ARI dist. & 0.22 & 0.37 & 0.32 & 0.33 & 0.22 & 0.32 & 0.33 & 0.20 & \blue{0.16} & \red{0.00} \\
			all ARI dist. & 0.27 & 0.43 & 0.37 & 0.38 & 0.27 & 0.37 & 0.38 & 0.24 & \blue{0.2} & \red{0.00} \\
			test NMI dist. & 0.11 & 0.53 & 0.39 & 0.39 & 0.14 & 0.39 & 0.40 & 0.12 & \blue{0.09} & \red{0.00} \\
			all NMI dist. & 0.17 & 0.44 & 0.35 & 0.36 & 0.19 & 0.35 & 0.35 & 0.12 & \blue{0.11} & \red{0.00} \\
	\midrule
	test ARI & 0.18 & 0.03 & 0.08 & 0.07 & 0.17 & 0.08 & 0.07 & 0.19 & \blue{0.24} & \red{0.40} \\
			all ARI & 0.19 & 0.03 & 0.09 & 0.08 & 0.19 & 0.09 & 0.08 & 0.22 & \blue{0.26} & \red{0.46} \\
			test NMI & 0.54 & 0.12 & 0.26 & 0.26 & 0.51 & 0.26 & 0.25 & 0.53 & \blue{0.56} & \red{0.65} \\
			all NMI & 0.38 & 0.11 & 0.20 & 0.19 & 0.36 & 0.20 & 0.19 & 0.42 & \blue{0.44} & \red{0.55} \\
  \bottomrule
\end{tabular}
\end{table*}
We conclude that {\SSS} attains the best performance in terms of both ARI and NMI, with regards to both test nodes and all nodes, in each of the 21 years.
\vspace{-3mm}
%%%%%%%%%%%%%%%%%%%%%%
\subsubsection{GICS Alignments Plots on S\&P1500}
\begin{figure}[hbt!]
\vspace{-2mm}
\centering   
\includegraphics[width=0.8\linewidth, trim=0.6cm 0.7cm 0.5cm 0.5cm,clip]{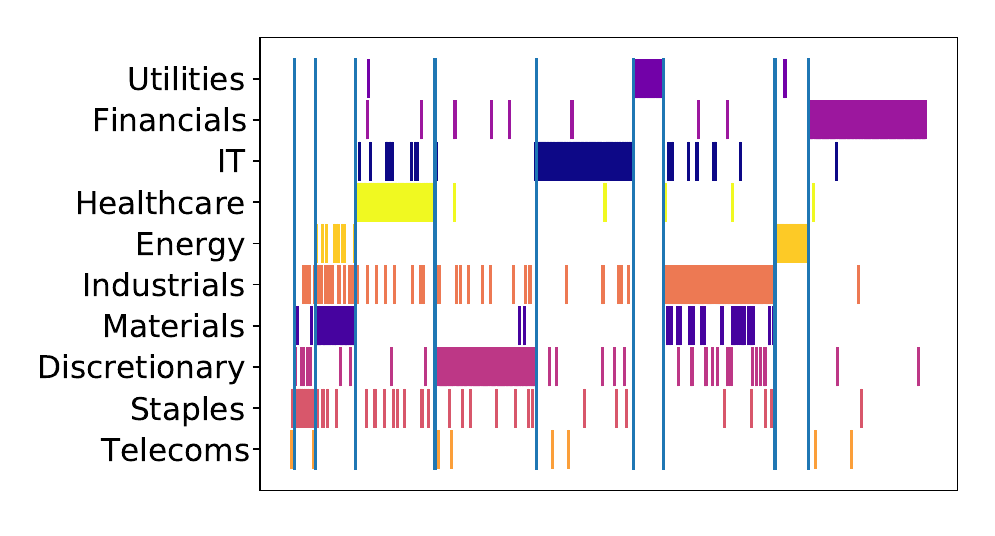}
\vspace{-2mm}
\caption{Alignment of  {\SSS} clusters with GICS 
sectors in S\&P 1500; ARI=0.71. Colors denote distinct sectors of the US economy, indexing  the rows; the total area of a color denotes the size of a GICS sector. Columns index the  recovered {\SSS} clusters, with the widths proportional to cluster sizes.
}
\vspace{-2mm}
\label{SSSNET_fig:SP1500ind}
\vspace{-2mm}
\end{figure}
We remark that {\SSS} (Figure \ref{SSSNET_fig:SP1500ind}) uncovers several very cohesive clusters, such as IT, Discretionary, Utility, and Financials. These recovered clusters are visually more cohesive than those reported by SPONGE. 
%%%%%%%%%%%%%%%%%%%%%%%%%%%%%%%%%%%%%%%%%%%%%%%%%%%%
% real-world data description
%%%%%%%%%%%%%%%%%%%%%%%%%%%%%%%%%%%%%%%%%%%%%%%%%%%%%%
\vspace{-3mm}
\section{Extended Data Description}
\label{SSSNET_appendix:data_description_more}
\subsection{SSBM Construction Details}
\label{SSSNET_appendix:SSBM_more}
A Signed Stochastic Block Model (SSBM) for a network on $n$ nodes  with $K$ blocks (clusters), is constructed   similar to \cite{SPONGE_AISTATS_2019} but with a more general cluster size definition.

\noindent  \bb (1) Assign block sizes $n_0 \le n_1 \le \cdots \le n_{K-1}$ 
with size ratio $\rho \geq 1$, as follows. If $\rho=1$, then the first $K-1$  blocks have the same size $\lfloor n/K \rfloor$,  and the last block has size $ n - (K-1) \lfloor n/K \rfloor$. If $\rho > 1$, we set 
$\rho_0=\rho^{\frac{1}{K-1}}.$
Solving 
$\sum_{i=0}^{K-1}\rho_0^in_0=n$ and taking
integer value gives $n_0=\left\lfloor {n(1-\rho_0)} / {(1-\rho_0^K)}\right\rfloor.$
Further, set $n_i=\lfloor\rho_0n_{i-1}\rfloor,$ for $i=1,\cdots, K-2$ if $K\geq 3,$ and $n_{K-1}=n-\sum_{i=0}^{K-2}n_i.$
Then, the ratio of the size of the largest to the smallest block is approximately $\rho_0^{K-1}=\rho.$ 
\noindent\bb   (2) 
Assign each node to one of $K$ blocks, so that each block has the allocated size. 
\noindent\bb  (3)  
For each pair of nodes in the same block, with probability $p_\text{in}=p$,  create an edge with $+1$ as weight between them, independently of the other potential edges. 
\noindent\bb (4) 
For each pair of nodes in different blocks,  with probability $p_\text{out}=p,$  create an edge with $-1$ as weight between them, independently of the other potential edges. 
\noindent\bb (5) 
Flip the sign of the across-cluster edges from the previous stage with sign flip probability $\eta_\text{in}=\eta$,  and $\eta_\text{out}=\eta$ for edges within and across clusters, respectively.

As our framework can be applied to different connected components separately, after generating the initial SSBM, we concentrate on 
the largest connected component.
To further avoid numerical issues, we modify  
the synthetic network by adding randomly wired edges to nodes of degree 1 or 2. 

The actual implementation of the above algorithm is given in \url{https://github.com/SherylHYX/SSSNET_Signed_Clustering/blob/main/src/utils.py}, modified from \url{https://github.com/alan-turing-institute/SigNet/blob/master/signet/block_models.py}.
%%%%%%%%%%%%%%%%%%%%%%%%%%%%%%%%%%%%%%%%%%%%%%%
\vspace{-3mm}
\subsection{Discussion on \textsc{POL-SSBM}}
\label{SSSNET_appendix:polarized_discussion}
Our generalizations are as follows: 
\bb
(1)  Our model allows for different sizes of communities and blocks, governed by the parameter $\rho$, which is more realistic.  
\bb
(2) 
Instead of using a single parameter for the edge probability and sign flips, we use two parameters $p$ and $\eta$. We also assume equal edge sampling probability throughout the entire graph, as we want to avoid being able to trivially solve the problem by 
considering the absolute value of the edge weights, and thus falling back onto the standard community detection setting in unsigned graphs.
\bb
(3)  We consider more than two polarized communities,  while also allowing for the existence of ambient nodes in the graph, in the spirit of \cite{Xiao}. 
%%%%%%%%%%%%%%%%%%%%%%%%%%%%%%%%%%%%%%%%%%%%
\vspace{-3mm}
\subsection{Real-World Data Description}
\label{SSSNET_appendix:real_description}
We perform experiments on six real-world signed network data sets (\textit{Sampson} \cite{sampson1969novitiate},
\textit{Rainfall} \cite{bertolacci2019climate},
\textit{Fin-YNet},
\textit{S\&P 1500} \cite{SP1500},
\textit{PPI} \cite{vinayagam2014integrating}, 
and \textit{Wiki-Rfa} \cite{west2014exploiting}).
Table~\ref{SSSNET_tab:data_sets} 
in the main text gives some summary statistics; here is a brief description of each data set.

\noindent 
\bb 
The Sampson monastery data \cite{sampson1969novitiate}
were collected by Sampson while resident at the monastery;
the study spans 12 months.  
This data set contains relationships (esteem, liking, influence, 
praise, as well as disesteem, negative influence, and blame) between  25 novices in total, who were preparing to join  
a New England monastery. Each novice was asked to rank their top three choices for each of these relationships. 
Some novices gave ties for some of the choices, and nominated four instead of three other novices. the positive attributes have values 1, 2 and 3 in increasing order of affection, whereas the negative attributes take values -1, -2, and -3, in increasing order of dislike. The social 
relations were measured at five points in time. Some novices had left the monastery during this process; at time point 4, 18 novices are present. These novices possess as feature whether or not they attended the minor seminary of `Cloisterville' before coming to the monastery. For the other 7 novices this information is not available. 
We combine these relationships into a network of 25  nodes by adding the weights for each relationship across all time points. Missing observations were set to 0.
Based on his observations and analyses, Sampson divided the novices into four groups: Young Turks, Loyal Opposition, Outcasts, and an interstitial group; this division is taken as ground truth. We use as node (novice) attribute 
whether or not they attended `Cloisterville'  before coming to the monastery. 

\noindent
\bb
\textit{Rainfall} \cite{bertolacci2019climate} contains 64,408 pairwise correlations between $n=306$ locations in Australia, 
at which historical rainfalls have been measured.

\noindent
\bb
\textit{Fin-YNet} consists of yearly  correlation matrices for $n=451$ stocks for 2000-2020 (21 distinct networks), using \textit{market excess returns}. That is, we compute each correlation matrix from overnight (previous close to open) and intraday (open-to-close) price daily returns, from which we subtract the market return of  the S\&P500 index for the same time interval.  In other words, within a given year, for each stock, we consider the time series of 500 market excess returns (there are 250 trading days within a year, and each day contributes with two returns, an overnight one and in intraday one). Each correlation network is built from the empirical correlation matrix of the entire set of stocks. For this data set, we report the results averaged over the 21 networks.

\noindent 
\bb 
\textit{S\&P1500} \cite{SP1500} considers daily prices for $n=1,193$ stocks in the S\&P 1500 Index, between 2003 and 2015, and builds correlation matrices from market excess returns (ie, from the return price of each financial instrument, the return of the market S\&P500 is subtracted). Since we do not threshold, the result is thus a fully-connected weighted network, with stocks as nodes and correlations as edge weights.

\noindent
\bb
\textit{PPI} \cite{vinayagam2014integrating} is a signed protein-protein interaction network between $n=3,058$ proteins.

\noindent 
\bb 
\textit{Wiki-Rfa} \cite{west2014exploiting} is a signed network describing voting information for electing Wikipedia managers. Positive edges represent supporting votes, while negative edges represent opposing votes. We extract the largest connected component and remove nodes with degree at most one, resulting in $n=7,634$ nodes for experiments.

%%%%%%%%%%%%%%%%%%%%%%%%%%%%%%%%%%%%%%%%%%%%%%%%%%%%
% implementation details
%%%%%%%%%%%%%%%%%%%%%%%%%%%%%%%%%%%%%%%%%%%%%%%%%%%%%%
\section{Implementation Details}
\label{SSSNET_appendix:implementation}
\subsection{Efficient Algorithm for SIMPA}
\label{appendix_subsec:simpa}
An efficient implementation of SIMPA is given in Algorithm \ref{algo:SIMPA}.
We omit the subscript $\mathcal{V}$ for ease of notation.
The matrix operations described in Eq.~\eqref{SSSNET_eq:g_sp} 
in the main text appear to be computationally expensive and space unfriendly. However, {\SSS} resolves these concerns via an efficient sparsity-aware implementation without explicitly calculating the sets of powers, such as $\mathcal{A}^{s+,h}.$ The algorithm also takes sparse matrices as input, and sparsity is maintained throughout.  
Therefore, for input feature dimension $d_\text{in}$ and hidden dimension $d$, if $d'=\max(d_\text{in},d) \ll n,$ time and space complexity of SIMPA, and implicitly {\SSS}, is  $\mathcal{O}(|\mathcal{E}|d'h^2+4nd'K)$ and $\mathcal{O}(4|\mathcal{E}|+10nd'+nK),$ respectively \cite{harrison2018high,complexity}.
When the network is large, SIMPA is amendable to a minibatch version using neighborhood sampling, similar to the minibatch forward propagation algorithm in \cite{Hamilton2017, GTTF2021}.
SIMPA is also amenable to an auto-scale version with theoretical guarantees, following \cite{Fey/etal/2021}.
\vspace{-3mm}
\begin{algorithm2e}
\SetAlgoLined
\SetKwInOut{Input}{Input}
\SetKwInOut{Output}{Output}
\SetKw{KwBy}{by}
\Input{(Sparse) row-normalized adjacency matrices $\overline{\mathbf{A}}^{s+},\overline{\mathbf{A}}^{s-},\overline{\mathbf{A}}^{t+},\overline{\mathbf{A}}^{t-}$; initial hidden representations $\mathbf{H}^{s+},\mathbf{H}^{s-},\mathbf{H}^{t+},\mathbf{H}^{t-}$; hop $h$; lists of scalar weights $\Omega^{s+} =( \omega_\mathbf{M}^{s+}, \mathbf{M} \in \mathcal{A}^{s+,h}),
\Omega^{s-} =( \omega_\mathbf{M}^{s-}, \mathbf{M} \in \mathcal{A}^{s-,h}),$
$\Omega^{t+}=( \omega_\mathbf{M}^{t+}, \mathbf{M} \in \mathcal{A}^{t+,h}),
\Omega^{t-}=( \omega_\mathbf{M}^{t-}, \mathbf{M} \in \mathcal{A}^{t-,h}).$ }
\Output{Vector representations $\mathbf{z}_i$ for all $v_i\in \mathcal{V}$ given by $\mathbf{Z}.$}
$\mathbf{Z}^{s+} \leftarrow \Omega^{s+}[0]\cdot \mathbf{H}^{s+}$;
$\mathbf{Z}^{t+} \leftarrow \Omega^{t+}[0]\cdot \mathbf{H}^{t+}$;
 $\mathbf{Z}^{s-}, \mathbf{Z}^{t-} \leftarrow \mathbf{0}$\;
 $\Tilde{\mathbf{X}}^{s+} \leftarrow \mathbf{H}^{s+},\bar{\mathbf{X}}^{s-} \leftarrow \mathbf{H}^{s-},\Tilde{\mathbf{X}}^{t+} \leftarrow \mathbf{H}^{t+},\bar{\mathbf{X}}^{t-} \leftarrow \mathbf{H}^{t-}$;
 $j \leftarrow 0$\;
  \For{$i\gets0$ \KwTo $h$}{
    \If{$i>0$}{
    $\Tilde{\mathbf{X}}^{s+} \leftarrow \overline{\mathbf{A}}^{s+}\Tilde{\mathbf{X}}^{s+}$;
    $\Tilde{\mathbf{X}}^{t+} \leftarrow \overline{\mathbf{A}}^{t+}\Tilde{\mathbf{X}}^{t+}$
    \;
    $\mathbf{Z}^{s+} \leftarrow \mathbf{Z}^{s+}+\Omega^{s+}[i]\cdot \Tilde{\mathbf{X}}^{s+}$;
    $\bar{\mathbf{X}}^{s-} \leftarrow \overline{\mathbf{A}}^{s+}\bar{\mathbf{X}}^{s-}$\;
    $\mathbf{Z}^{t+} \leftarrow \mathbf{Z}^{t+} + \Omega^{t+}[i]\cdot \Tilde{\mathbf{X}}^{t+}$;
    $\bar{\mathbf{X}}^{t-} \leftarrow \overline{\mathbf{A}}^{t+}\bar{\mathbf{X}}^{t-}$;
    }
    \If{$i\neq h$}{
    $\Tilde{\mathbf{X}}^{s-} \leftarrow \overline{\mathbf{A}}^{s-}\bar{\mathbf{X}}^{s-}$;
    $\Tilde{\mathbf{X}}^{t-} \leftarrow \overline{\mathbf{A}}^{t-}\Tilde{\mathbf{X}}^{t-}$
    \;
    $\mathbf{Z}^{s-} \leftarrow \mathbf{Z}^{s-}+\Omega^{s-}[j]\cdot \Tilde{\mathbf{X}}^{s-}$\;
    $\mathbf{Z}^{t-} \leftarrow \mathbf{Z}^{t-} + \Omega^{t-}[j]\cdot \Tilde{\mathbf{X}}^{t-}$;
    $j \leftarrow j+1$\;
    \For{$k\gets0$ \KwTo $h-i-2$}{
            $\Tilde{\mathbf{X}}^{s-} \leftarrow \overline{\mathbf{A}}^{s+}\bar{\mathbf{X}}^{s-}$;
            $\Tilde{\mathbf{X}}^{t-} \leftarrow \overline{\mathbf{A}}^{t+}\Tilde{\mathbf{X}}^{t-}$
            \;
            $\mathbf{Z}^{s-} \leftarrow \mathbf{Z}^{s-}+\Omega^{s-}[j]\cdot \Tilde{\mathbf{X}}^{s-}$\;
            $\mathbf{Z}^{t-} \leftarrow \mathbf{Z}^{t-} + \Omega^{t-}[j]\cdot \Tilde{\mathbf{X}}^{t-}$;
            $j \leftarrow j+1$\;
        }
    }
    }
$\mathbf{Z}=\operatorname{CONCAT}\left(\mathbf{Z}^{s+},\mathbf{Z}^{s-},\mathbf{Z}^{t+},\mathbf{Z}^{t-}\right) $\;
 \caption{Signed Mixed-Path Aggregation (SIMPA) algorithm for signed directed networks}
 \label{algo:SIMPA}
\end{algorithm2e}

%%%%%%%%%%%%%%%%%%%%%%%%%%%%%%%%%%%%%%%%%%%%
\subsection{Machines}
Experiments were conducted on a compute node with 4 Nvidia RTX 8000, 48 Intel Xeon Silver 4116 CPUs and $1000$GB RAM, a compute node with 3 NVIDIA GeForce RTX 2080, 32 Intel Xeon E5-2690 v3 CPUs and $64$GB RAM, a compute node  with 2 NVIDIA Tesla K80, 16 Intel Xeon E5-2690 CPUs and $252$GB RAM, and an Intel 2.90GHz i7-10700 processor with 8 cores and 16 threads. 
With the above, most experiments can be completed within a day.
%%%%%%%%%%%%%%%%%%%%%%%%%%%%%%%%%%%%%%%%%%%%%%%%
\subsection{Data Splits and Input}
For each setting of synthetic data and real-world data, we first generate five different networks, each with two different data splits, then conduct experiments on them and report average performance over these 10 runs.

For synthetic data, 10\% of all nodes are selected as test nodes for each cluster (the actual number is the ceiling of the total number of nodes times 0.1, so we would not fall below 10\% of test nodes), 10\% are selected as validation nodes (for model selection and early-stopping; again, we take the ceiling for the actual number), while the remaining roughly 80\% are selected as training nodes (the actual number is bounded above by 80\% since we take ceiling). 
For most real-world data sets, we extract the largest weak connected component for experiments.
For Wiki-Rfa, we further rule out nodes that have degree less than two.

As for input features, we weigh the unit-length eigenvectors of the Signed Laplacian or regularized adjacency matrix by their eigenvalues introduced in \cite{hou2003laplacian}.
For the Signed Laplacian features, we divide each eigenvector by its corresponding eigenvalue, since smaller eigenvectors are more likely to be informative.
For regularized adjacency matrix features, we multiply eigenvalues by eigenvectors, since larger eigenvectors are more likely to be informative.
After this scaling, there are no further  standardization steps before inputting the features to our model.
When features are available  (in the case of \textit{Sampson} data set), we standardize the one-dimensional binary input feature, so that the whole vector has mean zero and variance one.

For the sns and dns methods defined in Sec.~\ref{SSSNET_subsec:results} 
in the main text, we stack the eigenvectors associated with the smallest $K$ eigenvalues of the corresponding Laplacians \cite{zheng2015spectral} to construct the feature matrix, then apply K-means to obtain the cluster assignments.
For the other implementations, we also take the first $K$ eigenvectors, either smallest or largest, following \url{https://github.com/alan-turing-institute/SigNet/blob/master/signet/cluster.py}.
%%%%%%%%%%%%%%%%%%%%%%%%%%%%%%%%%%%%%%%%%%%%%%%%%%%
\subsection{Hyperparameters}
\label{SSSNET_appendix:hyperparameters}
We conduct hyperparmeter selection via a greedy search manner. To explain the details, consider for example the following synthetic data setting:
polarized SSBMs with 1050 nodes, $n_c=2$ SSBM communities, $\rho=1.5, N=200, p=0.1.$

Recall that the the
objective function
minimizes  
\begin{equation}
    \mathcal{L} = \mathcal{L}_\text{PBNC}+\gamma_s( \mathcal{L}_\text{CE}+\gamma_t\mathcal{L}_\text{triplet}), 
\end{equation}
where $\gamma_s,\gamma_t>0$ are weights for the supervised part of the loss and triplet loss within the supervised part, respectively.

Note that the cosine similarity (used in triplet loss $\mathcal{L}_\text{triplet}$) between two randomly picked vectors in $d$ dimensions is bounded by $\sqrt{ \ln (d)/d}$ with high probability. In our experiments $d=32$, and $\sqrt{ \ln (2d)/(2d)} \approx 0.25,\sqrt{ \ln (4d)/(4d)} \approx 0.19.$ In contrast, for fairly uniform clustering,  the cross-entropy loss grows like $\log n$, which in our experiments ranges between 3 and 17. Thus some balancing of the contribution is required.

Instead of a grid search, we tune hyperparameters according to what performs the best in  the current default setting. If two settings give similar results, we  pick the simpler setting, for example, the smaller hop size or the lower number of seed nodes.

When we reach a local optimum, we stop searching. Indeed, just a few iterations (less than five) were required for us to find the current setting, as {\SSS} tends to be robust to most hyperparameters.

\begin{figure}[ht]
    \centering
\begin{subfigure}[ht]{0.32\linewidth}
  \centering
      \includegraphics[width=\linewidth]{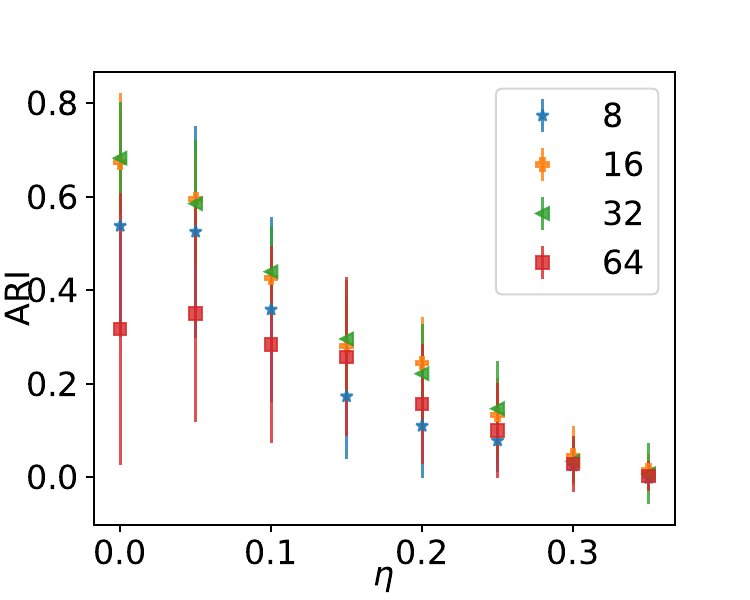}
      \caption{Vary $d$, the MLP hidden dimension}.
      \label{SSSNET_fig:polarized_hidden}
\end{subfigure}
    \begin{subfigure}[ht]{0.32\linewidth}
      \centering
      \includegraphics[width=\linewidth]{figures/ablation/test0_200_50_10_10_3200_0_0_5000Change_hop.pdf}
      \caption{Vary $h$, the number of hops for aggregation}.
      \label{SSSNET_fig:polarized_hop}
    \end{subfigure}
    \begin{subfigure}[ht]{0.32\linewidth}
      \centering
      \includegraphics[width=\linewidth]{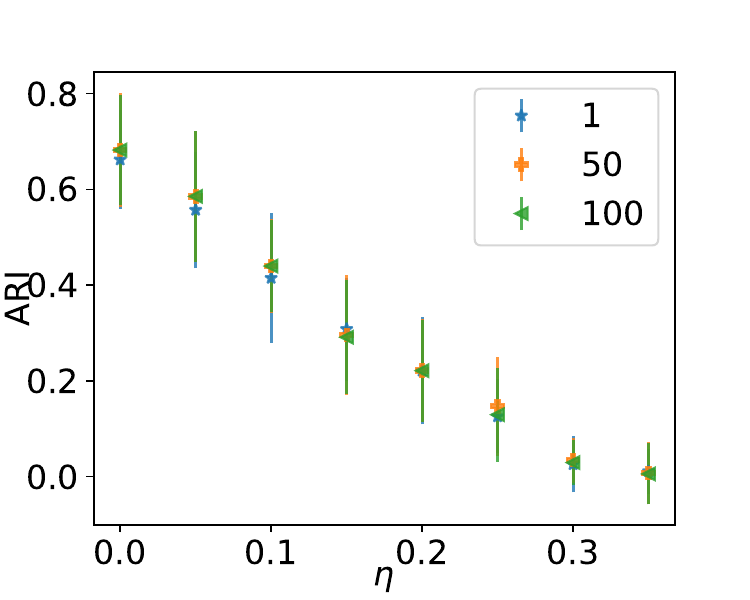}
      \caption{Vary $\tau$, weight of the added self-loop to $\mathbf{A}^+$}.
      \label{SSSNET_fig:polarized_tau}
    \end{subfigure}
    \begin{subfigure}[ht]{0.32\linewidth}
  \centering
  \includegraphics[width=\linewidth]{figures/ablation/test0_200_50_10_10_3200_0_0_5000Change_supervised_loss_ratio.pdf}
  \caption{Vary $\gamma_s$ in Eq.~\eqref{SSSNET_eq:loss_overall} in main.}
  \label{SSSNET_fig:polarized_gamma_s}
    \end{subfigure}
    \begin{subfigure}[ht]{0.32\linewidth}
      \centering
      \includegraphics[width=\linewidth]{figures/ablation/test0_200_50_10_10_3200_0_0_5000Change_triplet_loss_ratio.pdf}
      \caption{Vary $\gamma_t$ in Eq.~\eqref{SSSNET_eq:loss_overall} in main.}
    \end{subfigure}
    \begin{subfigure}[ht]{0.32\linewidth}
      \centering
      \includegraphics[width=\linewidth]{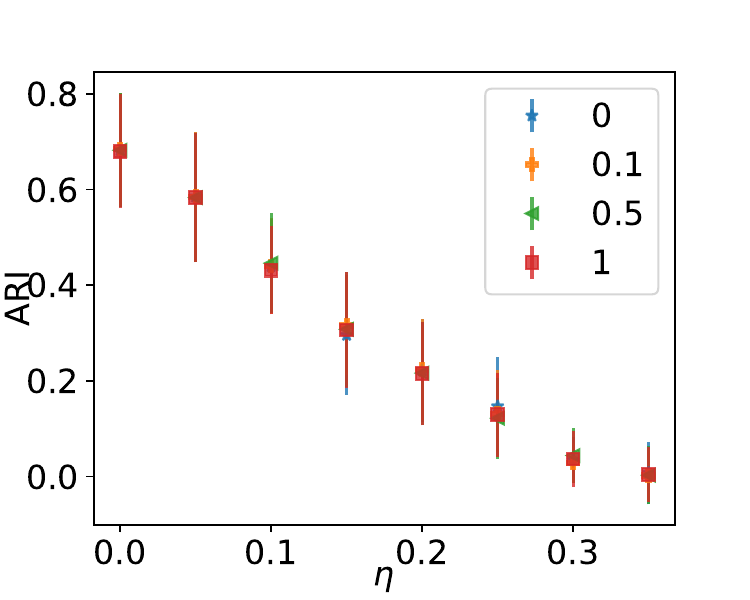}
      \caption{Vary $\alpha$ in $\mathcal{L}_\text{triplet}$.}
      \label{SSSNET_fig:polarized_alpha}
    \end{subfigure}
    \vspace{-3mm}
    \caption{Hyperpameter analysis on polarized SSBMs with $n=1050$ nodes, $n_c=2$ communites, $\rho=1.5,$ default community size $N=200$, and $p=0.1.$  
    } 
    \label{SSSNET_fig:hyper}
\end{figure}

Figure \ref{SSSNET_fig:hyper} compares the performance of {\SSS} on polarized SSBMs with 1050 nodes, $n_c=2$ SSBM communities, $\rho=1.5, N=200, p=0.1,$ under different hyperparameter settings.
By default, we use the loss function Eq.~\eqref{SSSNET_eq:loss_overall} in the main text 
with $\gamma_t=0.1, \gamma_s=50,$ and $d=32,l=2,\tau=0.5,h=2, \alpha=0.$
We use the default seed ratio as 0.1 (the ratio of the number of seed nodes to the number of training nodes). 
We remark from (a) that as we increase the MLP hidden dimension $d$, performance first improves then decreases, with 32 a desirable value. 
As we increase the number of hops to consider, performance drops in (b). Therefore, we would use the simplest, yet best choice, hop 2.
The decrease in both cases might be explained by too much noise introduced.
For (c), the self-loop weight $\tau$ added to the positive part of the adjacency matrix does not seem to affect performance much, and hence we use 0.5 throughout. 
We conclude from (d) that it is recommended to have $\gamma_s >1$ so as to take advantage of labeling information. 
The best triplet loss ratio in our candidates is $\gamma_t=0.1$, based on (e). 
From (f), the influence of $\alpha$ is not evident, and hence we use $\alpha=0$ throughout.
%%%%%%%%%%%%%%%%%%%%%%%%%%%%%%%%%%%%%%%%%%%%%%%%
\subsection{Training}
For all synthetic data, we train {\SSS} with a maximum of 300 epochs, and stop training when no gain in validation performance is achieved for 100 epochs (early-stopping).

For real-world data, when ``ground-truth" labels are available, we still have separate test nodes.
For S\&P1500 data set, we do not have validation nodes, and 90\% of all nodes are training nodes.
For data sets with no ``ground-truth" labels available, we train {\SSS} in a self-supervised setting, using all nodes to train. We stop training when the training loss does not decrease for 100 epochs or when we reach the maximum number of epochs, 300.

For the two-layer MLP, we do not have a bias term for each layer, and we use a Rectified Linear Unit (ReLU),  followed by a dropout layer with 0.5 dropout probability between the two layers, following \cite{Tian}. 
We use Adam as the optimizer, and $ \ell_2 $ regularization with weight decay $5\cdot 10^{-4}$ to avoid overfitting.

%%%%%%%%%%%%%%%%%%%%%%%%%%%%%%%%
\subsection{Aggregation Matrices}
In practice, in order not to consider $h-$hop neighbors that are ``fake" (due to added self-loops), after obtaining the aggregation matrix in Sec.~\ref{sec:SIMPA}, for example, a matrix from $\mathcal{A}^{s-,h}=
     \left\{(\mathbf{\overline{A}}^{s+})^{h_1}\cdot\mathbf{\overline{A}}^{s-}\cdot(\mathbf{\overline{A}}^{s+})^{h_2}:h_1,h_2\in H 
   \right\} $, we set the entries that should not be nonzero if we do not add self-loops to zero. In other words, we only keep entries that exist in $(\mathbf{{A}}^{+})^{h_1}\cdot\mathbf{A}^{-}\cdot(\mathbf{A}^{+})^{h_2}$ for $h_1, h_2\in H.$ This trick originates from the implementation of \cite{Tian}.
\end{document}